\documentclass[aps,amsmath, amssymb,12pt,letterpaper]{article}
\usepackage{jheppub}

\usepackage{graphicx}
\usepackage{bm}
\usepackage{amssymb}
\usepackage{amsmath}
\usepackage{cancel}
\usepackage{color}

\newcommand{\pref}[1]{(\ref{#1})}

\def\bdm{\begin{displaymath}}
\def\edm{\end{displaymath}}

\newcommand{\dX}{{\mathsf X}}

\newcommand{\Back}{B\"{a}cklund }
\newcommand{\Sch}{Schr\"{o}dinger }
\newcommand{\Psit}{\tilde{\Psi}}

\newcommand{\xc}{x_{\mathrm{cl}}}
\newcommand{\xdc}{\partial_{t}x_{\mathrm{cl}}}
\newcommand{\ha}{\hat{a}}
\newcommand{\had}{\hat{a}^{\dagger}}
\newcommand{\pertsize}{\boldsymbol{a}}
\newcommand{\tsing}[1]{t_{\infty}(#1)}
\newcommand{\phichange}{\Delta\phi}

\newcommand{\Achange}{\Delta A}
\newcommand{\Bchange}{\Delta B}

\newcommand{\bra}[1]{\langle#1|}
\newcommand{\ket}[1]{|#1 \rangle}

\def\coeff#1#2{{\textstyle \frac #1 #2}}

\def\({\left(}
\def\){\right)}
\def\[{\left[}
\def\]{\right]}

\def\ie{{i.e.}}
\def\eg{{e.g.}}

\def\etc{{etc}}

\def\barray{\begin{array}}
\def\earray{\end{array}}
\def\be{\begin{equation}}
\def\ee{\end{equation}}
\def\ben{\begin{equation} \nonumber}
\def\een{\end{equation}}
\def\ban{\begin{eqnarray*}}
\def\ean{\end{eqnarray*}}
\def\bea{\begin{eqnarray}}
\def\eea{\end{eqnarray}}
\def\eal{\end{align}}
\def\bal{\begin{align}}
\def\nn{\nonumber}

\def\({\left(}
\def\){\right)}
\def\half{{1\over2}}

\def\One{{\hbox{ 1\kern-.8mm l}}}

\def\DD{{\mathcal{D}}}

\def\GG{{\mathcal{G}}}
\def\HH{{\mathcal{H}}}

\def\NN{{\mathcal{N}}}
\def\PP{{\mathcal{P}}}
\def\QQ{{\mathcal{Q}}}
\def\RR{{\mathcal{R}}}
\def\SS{{\mathcal{S}}}

\def\VV{{\mathcal{V}}}

\def\ghat{\hat g}
\def\aa{{\bf a}}

\def\R{R}

\def\atil{\alpha}
\def\btil{\beta}
\def\phitil{\tilde\phi}
\def\Xtil{\tilde X}
\def\mpl{m_{ p} }
\def\mpleff{m_{p,\rm eff} }

\def\cl{{\rm cl}}
\def\qu{{\rm qu}}
\def\eff{{\rm eff}}

\def\grav{{\rm grav}}
\def\matt{{\rm matt}}

\def\inflaton{X}		% inflaton
\def\infb{\hat{X}}	% inflaton background
\def\infpert{\bm x}	% inflaton perturbation
\def\infsub{{\scriptscriptstyle X}}	%inflaton subscript
\def\dX4d{{\delta \!X}}	% 4d inflaton perturbation

\def\Laux{\chi}			%Liouville aux field
\def\Lauxb{\hat\chi}		%Liouville aux field bkgd
\def\Lauxpert{{\bm \chi}}	%Liouville aux field pert

\def\phican{\phi}		%Liouville field 
\def\phiperp{\xi}	 	%modified Liouville theory field space direction orthogonal to phi
\def\phib{{\hat{\phican}}}	%Liouville field backgd
\def\phipert{\bm\varphi}	%Liouville field pert

\def\ntb{\hat N^t}		%lapse backgd
\def\nxb{\hat N^x}	%shift backgd
\def\ntpert{{{\bm n}^t}}	%lapse pert
\def\nxpert{{{\bm n}^x}}	%shift pert

\def\slsup{(\infb'/\phib')}%slow-roll suppression

\def\boost{\lambda}

\def\dS{de~Sitter }

\title{Modeling Quantum Gravity Effects in Inflation}

\author[ ]{Emil J. Martinec,}

\author[ ]{Wynton E. Moore}

\affiliation[ ]{Enrico Fermi Institute and Department of Physics, University of Chicago\\ 5640 S. Ellis Ave., Chicago, IL 60637-1433, USA}
\emailAdd{ejmartin@uchicago.edu} 
\emailAdd{wyntonmoore@uchicago.edu}

\abstract{
Cosmological models in 1+1 dimensions are an ideal setting for investigating the quantum structure of inflationary dynamics -- gravity is renormalizable, while there is room for spatial structure not present in the minisuperspace approximation.  We use this fortuitous convergence to investigate the mechanism of slow-roll eternal inflation.  A variant of 1+1 Liouville gravity coupled to matter is shown to model precisely the scalar sector of cosmological perturbations in 3+1 dimensions.  A particular example of quintessence in 1+1d is argued on the one hand to exhibit slow-roll eternal inflation according to standard criteria; on the other hand, a field redefinition relates the model to pure \dS gravity coupled to a free scalar matter field with no potential.  This and other examples show that the standard logic leading to slow-roll eternal inflation is not invariant under field redefinitions, thus raising concerns regarding its validity.  Aspects of the quantization of Liouville gravity as a model of quantum \dS space are also discussed.
  }

\begin{document}

\maketitle
\flushbottom

%%%%%%%%%%%%%%%%%%%%%%%%%%%%%%%%%%%%%%%%%%
%%%%%%%%%%%%%%%%%%%%%%%%%%%%%%%%%%%%%%%%%%

\section{Introduction}

\subsection{Two-dimensional gravity and cosmological models}

The analysis of quantum gravity effects in realistic, four-dimensional cosmological models is hampered by our present inability to quantize gravity in cosmological spacetimes.  In two spacetime dimensions, one does have a renormalizable theory of gravity -- Liouville field theory~\cite{Polyakov:1981rd} and variants thereof.  As we review below in section~\ref{sec:2dModels}, Liouville theory describes 2d gravity around a \dS or anti-deSitter background; coupling to matter, one finds two-dimensional versions of the Friedmann equations of cosmology~\cite{DaCunha:2003fm}.  Thus we have an ideal setting to investigate the structure of gravitational backreaction at the quantum level -- there is enough structure in the single spatial dimension to accommodate the inhomogeneous fluctuations that lead to structure formation, while gravity might be under sufficient control that we can hope to track the quantum back-reaction of the metric on the quantum matter fluctuations.

Of course, the gravity sector does not have independent field-theoretic degrees of freedom in 2d; there are no transverse traceless tensor fluctuations.  There is however a sector of scalar metric fluctuations, and as we show in section~\ref{sec:fluctuations} these behave precisely like their four-dimensional counterparts.  We develop a variant of Liouville theory for which there is a one-to-one correspondence between the scalar geometric perturbations in Liouville gravity and those of four-dimensional Einstein gravity.  The behavior of these fields under linearized gauge transformations is identified, and invariant combinations are constructed.  The quadratic effective action is written in terms of a precise analogue of the Mukhanov-Sasaki variable~\cite{Sasaki:1986hm,Mukhanov:1988jd}, which possesses linearized gauge invariance and exhibits a scale invariant fluctuation spectrum at this order.  Thus we have an ideal situation in which to model quantum gravity effects in inflationary cosmology~-- a renormalizable theory of gravity whose field content and perturbative structure matches that of the scalar sector of four-dimensional Einstein gravity.

%%%%%%%%%%%%%%%%%%%%%%%%%%%%%%%%%%%%%%%%%%
%%%%%%%%%%%%%%%%%%%%%%%%%%%%%%%%%%%%%%%%%%

\subsection{The idea of slow-roll eternal inflation}

A situation in which quantum effects play a key role is that of inflation.  The solutions to the classical equations of motion in an inflation model involve a scalar field, the inflaton $\inflaton$, slowly descending its smooth potential $\VV(\inflaton)$.  If the descent is slow enough, potential energy dominates over kinetic energy, and the matter equation of state approximates that of a cosmological constant.  The quantitative measure of ``slow enough'' is that the slow-roll parameters which measure the rate of variation of the Hubble scale $H$,
\bea
\label{slowrollgeneral}
\epsilon_H = -\frac{\dot H}{H^2} 
~,~~
\eta_H = \epsilon_H+\half\frac{\dot\epsilon_H}{H\epsilon_H}  ~,
\eea
are much smaller than one over the course of the inflationary epoch.

In the quantum theory of a scalar field rolling down its potential in curved spacetime, there are fluctuations about the classical field value.  These fluctuations back-react on the geometry to make curvature perturbations -- indeed, these curvature perturbations are thought to seed the fluctuations in the Cosmic Microwave Background that we observe today.

The idea of slow-roll eternal inflation is that large coherent fluctuations $\delta\inflaton$ over Hubble-size volumes, seemingly far out on the tails of the scalar field probability distribution, nevertheless have an extraordinary effect on the structure of the wavefunction\cite{Steinhardt:1982,Vilenkin:1983xq,Linde:1986fd,Goncharov:1987ir,Linde:1993xx}.  Consider what happens in a single Hubble volume $H^{-n}$ in $n$ spatial dimensions, over a Hubble time $H^{-1}$.  The classical motion shifts the inflaton from its initial value $\inflaton_i$ by $\delta\inflaton_\cl \approx \dot \inflaton_\cl H^{-1}$.  The wavefunction of the inflaton is centered on the classical value, with a width $|\delta\inflaton_\qu| \lesssim H/{2\pi}$~\cite{Linde:1982uu,Vilenkin:1982wt,Starobinsky:1982ee}.  The net shift of the inflaton
\be
\delta\inflaton = \delta\inflaton_\cl + \delta\inflaton_\qu
\ee
determines an effective cosmological constant $V(\inflaton_i + \delta\inflaton)$.  The amount of inflation in instances where $\delta\inflaton$ shifts the field to a higher value of the potential is greater than in instances where it shifts the field to a lower value of the potential.  If $\delta\inflaton_\qu$ is sufficiently larger than $\delta\inflaton_\cl$, then there is a substantial probability to make large volumes where the inflaton has fluctuated up, relative to the spatial volume generated by the classical motion.  Eventually, when one looks at large volume, the probability density is concentrated on field configurations where the inflaton has never rolled down its potential.
A quantitative analysis~\cite{Creminelli:2008es,Dubovsky:2011uy} indicates that a phase transition to this eternally inflating state occurs when
\be
\label{SREIcriterion}
\frac{\pi^{1+n/2}}{n\Gamma(n/2)}
\frac{\dot\inflaton^2}{H^{n+1}} \lesssim 1 ~.
\ee

%%%%%%%%%%%%%%%%%%%%%%%%%%%%%%%%%%%%%%%%%%
%%%%%%%%%%%%%%%%%%%%%%%%%%%%%%%%%%%%%%%%%%

\subsection{Quantum dS spacetime, and slow-roll eternal inflation in 2d}

To investigate the mechanism of slow-roll eternal inflation in the full quantum theory, in section~\ref {sec:ModelEquivalence} we focus on the particular choice of potential $\VV(\inflaton)=\exp[-\beta \inflaton]$ for a scalar matter field $\inflaton$ in a modified two-dimensional Liouville gravity.  For small $\beta$, this potential satisfies the criterion~\pref{SREIcriterion} for slow-roll eternal inflation.  On the other hand, a simple field redefinition relates this theory to pure \dS Liouville gravity and a decoupled free matter field $\tilde\inflaton$.  Tracked through the field redefinition, the prediction of slow-roll eternal inflation translates into a force pushing the field $\tilde\inflaton$ in a preferred direction, which on the face of it seems absurd.  In particular, in conformal gauge the field $\tilde\inflaton$ is completely decoupled.  The generalization from an exponential quintessence potential to a cosh potential exhibits similar inconsistencies.  Thus, to the extent that these models capture the structure of inflationary dynamics, it appears that the paradigm of slow-roll eternal inflation is logically inconsistent, since it is not invariant under field redefinitions.  Possible objections to the analysis are raised and discussed.

This result leads us to examine the quantization of \dS Liouville theory in section~\ref{sec:semiclassical}, using both WKB analysis as well as a relation to free field theory via the (canonical) \Back transformation.  We undertake this analysis in part to address some of the concerns raised in section~\ref {sec:ModelEquivalence}, but also because this theory is a model of two-dimensional quantum \dS spacetime and therefore of interest in its own right.  Classically, the \Back transformation is a canonical change of variables; at the quantum level, it is the functional equivalent of an integral transform~\cite{Braaten:1982fr,Braaten:1982yn,Moore:1991ag,Teschner:2001rv}, and allows one to quantize the dual free field, then construct the wavefunctional of the Liouville theory as an integral transform of the free field wavefunctional; the kernel of the integral transform is given by the generating functional of the canonical transformation.  We explore some properties of the resulting Liouville wavefunctional.  

%%%%%%%%%%%%%%%%%%%%%%%%%%%%%%%%%%%%%%%%%%
%%%%%%%%%%%%%%%%%%%%%%%%%%%%%%%%%%%%%%%%%%

\section{2d gravity and 2d inflationary cosmology}
\label{sec:2dModels}

\subsection{Timelike Liouville field theory}

Einstein gravity is trivial in two dimensions -- the Einstein-Hilbert action is a topological invariant (the Euler character of the two-dimensional spacetime); hence the Einstein tensor vanishes identically.  Liouville gravity~\cite{Polyakov:1981rd} provides a useful substitute, however~--~especially when it comes to cosmology~\cite{Polchinski:1989fn,DaCunha:2003fm}.  The Liouville action (for a review and further references, see~\cite{Ginsparg:1993is}; our conventions differ slightly so as to conform more closely to Einstein gravity in higher dimensions)
\be
\label{Sliou}
\SS_{L} = \frac{1}{2\pi \gamma^2} \int d^2\xi\sqrt{-\ghat}\Bigl[ -(\hat\nabla\phican)^2 - Q \hat\R[\ghat]\phican
- {\Lambda} e^{2\phican} \Bigr]
\ee
describes the dynamics of the scale factor $\phican$ of the (Lorentz signature) 2d metric in conformal gauge
\be
ds^2 = e^{2\phican} \ghat_{ab}d\xi^ad\xi^b ~.
\ee
Here $\ghat$ is a fixed background metric, whose scalar curvature is $\R[\ghat]$.
If one sets $Q=1$, the classical equation of motion
\be
\label{Leom}
\R[e^{2\phican}\ghat] = e^{2\phican}\bigl(-2\hat\nabla^2\phican+\R[\ghat]\bigr)
= - 2{\Lambda}
\ee
is solved by constant curvature `dynamical' metrics $g=e^{2\phican}\ghat$, \ie\ two-dimensional (anti)\dS spacetimes.
The parameter $\gamma^2$ plays the role of Newton's constant here; weak coupling is small $\gamma$.  In contrast to Einstein gravity in higher dimensions, the coupling $\gamma$ is dimensionless.

Quantum consistency of 2d gravity coupled to conformally invariant matter requires the vanishing of the total stress tensor
\be
\label{graveqs}
0 =  \langle T^{\rm tot}_{ab} \rangle = \langle T_{ab}^{\rm Liou} + T_{ab}^{\rm matter} + T_{ab}^{\rm ghost} \rangle
\ee
which includes contributions from the quantum Liouville field theory, conformal matter fields, and Faddeev-Popov ghosts for the coordinate gauge choice (for the moment, we work in conformal gauge where $\hat g$ is fixed).  In particular, the contributions to the conformal anomaly
\be
\label{anomaly}
\langle T_{a}^{~a} \rangle = -2\pi \ghat^{ab} \frac{\delta \SS_{\rm eff}}{\delta \ghat^{ab}} = \frac{c}{48\pi} \R[\ghat]
\ee
must cancel; this leads to a condition relating the various contributions to the conformal central charge $c$
\be
\label{ctot}
c_{\rm tot} \equiv c_{\rm L} + c_{\rm matt} + c_{\rm gh} = \left(1+12\frac{Q^2}{\gamma^2}\right) + c_{\rm matt} - 26 = 0
\ee
where the coefficient $Q$ appearing in the Liouville central charge receives a modification from its classical value $Q=1$ due to the quantization of the Liouville field itself,
\be
\label{Qval}
Q = 1 + \coeff12\gamma^2 ~.
\ee
This modification is determined by the condition of scale invariance of the exponential potential for the Liouville field at the semiclassical level.

Conformally invariant matter only couples to the Liouville field through the conformal anomaly and requirements of residual symmetry under conformal transformations, \ie\ only through the requirement~\pref{ctot} of vanishing of the total central charge, and through the stress tensor constraints~\pref{graveqs}.  In this case, the classical solution to the Liouville dynamics is a constant curvature dynamical metric $e^{2\phican} \ghat_{ab}$.  More generally -- either when the matter is non-conformal, or in gauges other than conformal gauge -- the gravitational and matter sectors interact non-trivially.

In Einstein gravity, the natural (de~Witt) metric on deformations $\delta g_{ab}$ in the metric configuration space has negative signature for deformations of the conformal factor $\delta\phican$.  This feature allows a nontrivial solution to the Hamiltonian constraint of the theory, for generic initial conditions in the classical theory; in cosmology, the timelike signature of the scale factor $\aa=e^{\phican}$ allows one to think of the scale factor as a measure of time in eras of uniform expansion or contraction.  To mimic this property in the 2d model, one wants the kinetic term of the Liouville field to similarly have a sign opposite to that of the matter fields.  This criterion is met for $c_{\rm matt}>25$, which leads to pure imaginary $\gamma$, and the desired opposite sign kinetic term for the metric conformal factor.   
In order to restore reality to the gravitational coupling, one maps 
%$\phican \to i\phican$, 
$\gamma\to -i\gamma$.
The action for this timelike Liouville theory reads
\be
\label{STL}
\SS_{\it TL} =  \frac{1}{2\pi\gamma^2} \int d^2\xi\sqrt{-\ghat}\Bigl[(\hat\nabla\phican)^2 +{Q} \hat\R[\ghat]\phican
-{\Lambda} e^{2\phican} \Bigr]
\ee
where now $Q=1 - \frac12\gamma^2$.  The requirement $c_{\rm matt} > 25$ can be satisfied by having a large number of matter fields; free scalars have $c=1$, free fermions $c=\half$.  In the language of inflationary cosmology, there is a large number of isocurvature modes.  Gravity is semiclassical in the limit of large $c_{\rm matt}$.

%%%%%%%%%%%%%%%%%%%%%%%%%%%%%%%%%%%%%%%%%%
%%%%%%%%%%%%%%%%%%%%%%%%%%%%%%%%%%%%%%%%%%

\subsection{More general models -- quintessence}

Models of inflationary cosmology arise if we allow a nontrivial potential for some of the matter fields, dressed by the dynamical metric~\cite{DaCunha:2003fm}.  A generic parity-symmetric action for scalar matter coupled to Liouville gravity is
\be
\label{sigmodel}
\SS  = \frac{1}{8\pi} \int d^2\xi \sqrt{-\ghat}\Bigl[ \ghat^{ab}\hat\nabla_a X^\mu \hat\nabla_b X^\nu \GG_{\mu\nu}(X) + \Phi(X) \R[\ghat] - \VV(X) \Bigr] ~,
\ee
where we account for the timelike Liouville dynamics via a Lorentz signature field space metric, together with the identification $X^0 = \phican$ among the $d+1$ fields $X^\mu$.  At the semiclassical level, the gravitational equations of motion imply the vanishing of the beta functions for the quantum field theory with this action; these are conditions on the coupling functions $\GG$, $\Phi$, and $\VV$~\cite{Friedan:1980jf,Friedan:1980jm,Fradkin:1984pq,Callan:1985ia,Sen:1985eb}
%\be
%\it beta~function~eqs
%\ee
At leading semiclassical order, one solution to these conditions is timelike Liouville theory coupled to $d>25$ free scalar matter fields; another solution which we will consider is the quintessence model
\be
\label{Squint}
\SS_{\it Q} =  \frac{1}{8\pi} \int d^2\xi\sqrt{-\ghat}\left[\frac{4}{\gamma^2}(\hat\nabla\phican)^2 -(\hat\nabla \inflaton)^2 
+ \hat\R[\ghat]\Bigl(\frac{4Q_\phi }{\gamma^2}\phican - Q_\infsub\inflaton\Bigr)
-{\mu^2} \exp\Bigl[2\alpha\phican - \beta\inflaton \Bigr]\right] 
\ee
together with $d-1$ additional, decoupled free matter fields.  The conditions of conformal invariance are
\be
-\frac{12 Q_\phi^2}{\gamma^2} + 3Q_x^2 + c_{\rm matt} - 24 = 0 ~~,\quad
-\bigl({\gamma^2}\alpha^2 + 2Q_\phi \alpha\bigr) + \bigl(\beta^2 + Q_\infsub\beta\bigr) + 2 = 0 ~. 
\ee
We will be interested in weakly varying matter potentials for which $\beta\ll\alpha$, and thus $\alpha\sim 1$, so that the gravitational dressing of the matter potential is close to the classical one.

Let us consider the equations of motion in the proper time coordinate $\tau$ instead of conformal time $t$,
\be
d\tau = e^{\phican} dt \equiv \aa dt ~,
\ee
and specialize to spatially homogeneous field configurations, and canonically scaling matter potential $\alpha\approx1$.  In terms of the `Hubble expansion rate'
\be
H = \frac{\dot \aa}{\aa} = \dot\phican
\ee
(here overdots denote proper time derivatives), and the matter potential $\VV(\inflaton) ={\mu^2} e^{-\beta\inflaton}$, the equations of motion and the Hamiltonian constraint then become
\bea
\label{friedmann}
0 &=& \ddot \inflaton + n H \dot\inflaton + {\VV_{,\infsub}} \nn\\
\dot H &=& -\frac{\gamma^2}{4}\[ \dot\inflaton^2 
+ \rho_\perp + P_\perp 
-\frac{1}{\aa^{2}}\(2+\frac14\Bigl(\frac{4Q_\phi^2}{\gamma^2}-Q_x^2\Bigr)\) \]   
\nn \\
H^2 &=& +\frac{\gamma^2}{4n}\[ \dot\inflaton^2 
+  \VV(\inflaton) + \rho_\perp 
-\frac{1}{\aa^{2}}\(2+\frac14\Bigl(\frac{4Q_\phi^2}{\gamma^2}-Q_x^2\Bigr)\) \]  ~. 
\eea
%\emil{get the Casimir energy right}
Here $n=1$ is the number of spatial dimensions, and $\rho_\perp$, $P_\perp$ are the spatially homogeneous energy density and pressure of the additional matter fields.  For $n>1$ spatial dimensions, these are precisely the Friedmann equations of Einstein gravity, if we identify $\gamma^2$ with the Newton constant $G_{\! N}$.  The $1/\aa^2$ terms in the last two equations are Casimir energy corrections to the stress tensor in a compact spatial geometry, and are absent if space is non-compact.
%; the term is similar to the spatial curvature contribution to the dynamics in higher dimensions.  
Thus the timelike Liouville/quintessence dynamics is very much the $d=2$ version of standard cosmological dynamics.

The conditions for slow-roll inflation are then that the dimensionless parameters
\bea
\label{slowroll}
\epsilon_H &=& -\frac{\dot H}{H^2} 
\approx \half \frac{2}{\gamma^2} \(\frac{\VV_{, \infsub}}{\VV} \)^2 \nn\\
\eta_H &=& \epsilon_H+\half\frac{\dot\epsilon_H}{H\epsilon_H}  
\approx   \frac{2}{\gamma^2}\[ \(\frac{\VV_{, \infsub \infsub}}{\VV}\)-\half\(\frac{\VV_{,\infsub}}{\VV}\)^2\]
\eea
are much smaller than one.  Slow-roll in the quintessence potential~\pref{Squint} thus holds provided $(\beta/\gamma)^2\ll 1$, which is simply the condition that the metric and matter potential have approximately their canonical scaling dimensions.

%%%%%%%%%%%%%%%%%%%%%%%%%%%%%%%%%%%%%%%%%%
%%%%%%%%%%%%%%%%%%%%%%%%%%%%%%%%%%%%%%%%%%

\subsection{Gauge-invariant formulation}

We will be interested in formulating cosmological perturbation theory in the above models, in a framework that allows comparison to results in four dimensions.  In particular we wish to begin from a gauge invariant starting point, rather than selecting conformal gauge at a very early stage of the analysis.  In the classical limit $Q=1$, a covariant generalization of the conformal anomaly term in the action~\pref{Sliou} is%
\be
\label{Lcovnonloc}
\int \sqrt g\R \, \frac1{\Delta} \sqrt g\R  ~,
\ee
where the spacetime curvature $\R$ and scalar Laplacian $\Delta$ are written in terms of a general metric $g_{ab}$.  Upon choosing conformal gauge $g_{ab}=e^{2\phican}\hat g_{ab}$, this action yields the Liouville kinetic term and background curvature coupling, up to a total derivative.  Its shortcoming is that it is nonlocal.  Therefore we introduce an auxiliary field $\Laux$ and write
\be
\label{Lcovpre}
\SS_\grav = \frac{1}{2\pi \gamma^2} \int \sqrt{-g}\left[  -(\nabla\Laux)^2 -  \R \Laux - {\Lambda} \right] ~,
\ee
%(dropping the coefficient $Q$ in front of the improvement term $\Laux \R$)
which yields~\pref{Lcovnonloc} upon eliminating $\Laux$ via its equation of motion.  Thus~\pref{Lcovpre} is a fully gauge invariant local action
which is classically equivalent to Liouville theory.  At the semiclassical level, this action receives quantum corrections and in conformal gauge becomes
\be
\label{Lcov}
\SS_\grav = \frac{1}{2\pi \gamma^2} \int \sqrt{-\hat g}\left[  -(\hat\nabla\Laux)^2 -  Q\hat\R \Laux  - 2\hat\nabla\Laux \cdot \hat\nabla\phican - {\Lambda}e^{2\phican} \right]  ~.
\ee
Because $\Laux$ is null direction in the field space, the $\hat R \chi$ term generates no net contribution to the central charge, and the condition~\pref{ctot} on the total central charge is $c_{\rm matt} = 24$, \ie\ as a string theory one has a light-like linear dilaton in the critical dimension.  This property is readily seen by diagonalizing the kinetic term; introducing $\xi=\chi+\phi$, one has
\be
\SS_\grav = \frac{1}{2\pi \gamma^2} \int \sqrt{-\hat g}\left[ (\hat\nabla\phican)^2 -(\hat\nabla\xi)^2 -  Q\hat\R (\xi-\phican )   - {\Lambda}e^{2\phican} \right]  ~,
\ee
\ie\ timelike Liouville theory together with a free matter field $\xi$ having a conformally improved stress tensor.
The conformal improvement terms of the spacelike $\xi$ and timelike $\phican$ contribute equal and opposite amounts to the central charge.

The scale dimension of the cosmological term $e^{2\phi}$ determines again $Q=1-\half\gamma^2$.  This value only arises after resumming self-contractions of the $\phican$ exponential, and one might expect the quantum corrections of the effective action to look different in other gauges.  In performing the covariant analysis of cosmological perturbations in the next section, we will consider the classical action and set $Q=1$, though in principle one could determine systematically the quantum corrections in a general gauge.

While on the subject of the covariance of 2d gravity, it is worth noting that an analysis of Liouville theory as a constrained Hamiltonian system was carried out by Teitelboim~\cite{Teitelboim:1983ux}.  The action takes the form
\be
\label{Ham}
\SS = \int \pi_\phican \partial_t\phican - N^t\HH - N^x \PP
\ee
where the Hamiltonian and momentum constraints are
\bea
\HH &=& \frac12\left( \kappa\pi_\phican^2 + \frac1\kappa (\partial_x\phican)^2\right)-\frac2\kappa\partial_x^2\phican - \frac\Lambda\kappa e^\phican
\\
\PP &=& \pi_\phican \partial_x\phican - 2 \partial_x\pi_\phican ~.
\eea
Naively one might think that this leads to a Lorentz covariant, gauge invariant theory when one allows the lapse and shift to be arbitrary, however upon passing to the Lagrangian formalism there are non-covariant terms involving the lapse and shift. Upon eliminating the field momentum from the action, one finds
\be
\SS = \frac{1}{2\kappa} \int \left[ \frac 1{N^t}\left(\partial_t \phican - N^x \partial_x\phican-2 \partial_x N^x\right)^2 
- N^t\left( (\partial_x\phican)^2 - 4\partial_x^2\phican - 2 \Lambda e^\phican\right)\right] ~.
\ee
This can be organized into the form~\pref{STL} with the background metric 
\be
\hat g_{ab} = \left(  \begin{matrix} -N_t^2+N_x^2\quad & N_x \\ N_x & 1 \end{matrix}\right) ~,
\ee
apart from a term $(\partial_x N^x)^2/N^t$.  This residual term has no Lorentz covariant expression.  Thus, while the Hamiltonian dynamics~\pref{Ham}, when coupled to matter, seems to have local reparametrization symmetry, it is not Lorentz invariant in a general gauge.  In particular, when performing the cosmological perturbation theory described in the next section around a slow-roll background, we have found that the action for small fluctuations is generically not Lorentz covariant.  Therefore, in what follows we will work with the covariant action~\pref{Lcov}, as well as its specialization to conformal gauge.

%%%%%%%%%%%%%%%%%%%%%%%%%%%%%%%%%%%%%%%%%%
%%%%%%%%%%%%%%%%%%%%%%%%%%%%%%%%%%%%%%%%%%

\section{Cosmological perturbation theory}
\label{sec:fluctuations}

A standard parametrization of the 2d metric is
\be
\label{metparam}
g_{ab} = e^{2\phican}\hat g_{ab} = e^{2\phican}\left(  \begin{matrix} -N_t^2+N_x^2\quad & N_x \\ N_x & 1 \end{matrix}\right) ~.
\ee
We expand around a spatially homogeneous background
\bea
\label{fieldperts}
\Laux &=& \Lauxb(t) + \Lauxpert(t,x) \nonumber\\
\phican  &=& \phib(t) + \phipert(t,x) \nonumber\\
N^t  &=& \ntb(t) + \ntpert(t,x) \\
N^x  &=& \nxb(t) + \nxpert(t,x) \nonumber\\
\inflaton  &=& \infb(t) + \infpert(t,x) \nonumber
\eea
and examine the structure of the perturbations following~\cite{Mukhanov:1988jd,Mukhanov:1990me} (for a recent review, see \eg~\cite{Baumann:2009ds}).  

%%%%%%%%%%%%%%%%%%%%%%%%%%%%%%%%%%%%%%%%%%
%%%%%%%%%%%%%%%%%%%%%%%%%%%%%%%%%%%%%%%%%%

\subsection{Background solutions}

For the purposes of this section, we will analyze the classical ($Q=1$) covariant 2d gravity theory~\pref{Lcov} coupled to classical matter, since that is the procedure followed in analyzing cosmological perturbations in higher dimensional Einstein gravity.
Working in conformal gauge (and in particular, conformal time) for the background, $\ntb=1$, $\nxb=0$, the background equations of motion are
\bea
0 &=& \phib'' + \Lauxb''  \nonumber\\
0 &=& \Lauxb'' + \Lambda e^{2\phib} + \frac{\gamma^2}{4}\alpha \,e^{2\alpha\phib} \VV(\infb)  \\
0 &=& -\infb'' + e^{2\alpha\phib}\VV_{,\infsub} (\infb)  \nonumber
\eea
(here prime denotes derivative with respect to conformal time); one also has the Hamiltonian constraint 
\be
0 = (\Lauxb')^2 + 2\Lauxb' \phib' + \Lambda e^{2\phib} + \frac{\gamma^2}{4}\left[ (\infb')^2 
 + e^{2\alpha\phib} \VV(\infb)\right] ~.
\ee

For instance, when $\VV=0$ one has deSitter solutions.
There are three general classes of such solutions for homogenous backgrounds:
\bea
\label{dsmetrics}
\Lambda \exp[{2\phican}] = 
\left\{
	\begin{array}{ll}
		&  \frac{\varepsilon^2}{\sinh^2(\varepsilon t)} \\
		& \frac{1}{t^2} \\
		& \frac{\varepsilon^2}{\sin^2(\varepsilon t)}
	\end{array}
\right.
\eea
where $t$ is conformal time.
Evaluating the stress-energy tensor of these solutions 
\bea
\label{Timproved}
T_{\pm\pm} = \frac{2}{\gamma^2}\[ (\partial_\pm\phi)^2 - \partial_\pm^2\phi \] = \HH\pm\PP 
\label{stresstens}
\eea
(in light-cone coordinates $x^\pm = t\pm x$, with compact spatial sections $x\sim x+2\pi $), 
one finds the Liouville field energy $E_L=(1+\varepsilon^2)/2\gamma^2$ 
for the first, `hyperbolic' solution;
$E_L= {1}/{2\gamma^2}$ 
for the second, `parabolic' solution; 
and 
$E_L = {(1-\varepsilon ^2)}/{2\gamma^2}$ 
for the third, `elliptic' solution.  These various solutions can be used to satisfy the stress tensor constraints~\pref{graveqs} for homogeneous states of the matter fields, depending on their energy.  The matter vacuum is paired with the global \dS solution, \ie\ the elliptic case~\pref{dsmetrics} with $\varepsilon =1$; increasing the matter energy decreases the Liouville momentum $\varepsilon $, pinching the neck of the \dS ``bounce'', until at the critical value $\varepsilon\to 0$ and the neck pinches off.  Increasing the matter energy further leads to the hyperbolic solutions, which have a Milne-type cosmological singularity as $t\to -\infty$.  The additive contribution ${1}/{2\gamma^2}$ in the stress tensor can be thought of as the Casimir energy of the matter fields on a spatial circle.  If we choose to work in a geometry with non-compact spatial sections (which is allowed for the parabolic and hyperbolic solutions), this Casimir term is absent, and the geometry is that of the parabolic solution.

%%%%%%%%%%%%%%%%%%%%%%%%%%%%%%%%%%%%%%%%%%
%%%%%%%%%%%%%%%%%%%%%%%%%%%%%%%%%%%%%%%%%%

\subsection{The quadratic fluctuation action}

We now wish to derive the effective action for perturbations to quadratic order~\cite{Mukhanov:1988jd,Mukhanov:1990me}.
Substituting the expansion~\pref{fieldperts} into the action~\pref{Lcov}, expanding to second order in $\epsilon$, and using the background equations of motion, one finds
\begin{multline}
\label{Squadeff}
\SS_2 = \frac{1}{2\pi\gamma^2 }\int \biggl( 
(\Lauxpert ')^2 -(\partial_x \Lauxpert )^2  +2 \phipert ' \Lauxpert ' -2 (\partial_x \phipert ) (\partial_x \Lauxpert )
\\
+\frac{\gamma^2 }{4}\Bigl[ (\infpert ')^2 -(\partial_x \infpert )^2  
-\frac{1}{2} e^{2\phib } \VV_{,\infsub\!\infsub}(\infb )\,\infpert^2
- 2 e^{2 \phib } \VV_{,\infsub}(\infb )\, \infpert \,\phipert \Bigr]  
\\
+ \Bigl[ \frac{\gamma^2 }{4}(\infb')^2  - (\phib ')^2\Bigr] \( (\ntpert)^2 +2 \ntpert \phipert  +2 \phipert^2 \) 
\\
 +2\partial_x \nxpert\Bigl[ \frac{\gamma^2 }{4}\infb ' \infpert  
 -2 \phib ' \phipert 
-2  \Lauxpert ' \Bigr] 
\\  
+2\ntpert \Bigl[  -   \phib ' \partial_x \nxpert + \partial_x^2\Lauxpert    
+ \phib '  \phipert ' 
- \frac{\gamma^2 }{4}\Bigl( \infb '  \infpert ' 
-\frac12e^{2 \phib } \VV_{,\infsub}(\infb ) \,\infpert \Bigr)  \Bigr]
 \biggr) ~,
\end{multline}
up to total derivatives. The equations of motion of the lapse and shift $\ntpert$ and $\nxpert$ yield constraint equations that can be solved for the lapse and shift, with the result
\bea
\ntpert &=& \frac{(\gamma^2 /4)\infb '\infpert -\phib '\phipert -\Lauxpert '}{\phib '} ~,
\\
\partial_x \nxpert &=& 
\( 1- \frac{\gamma^2 }{4}\slsup^2 \) \[ - \frac{\gamma^2 }{4}\infb '\infpert  +\Lauxpert'
\]
\nonumber\\
&&\qquad 
- \,\frac{\gamma^2 }{4} \slsup\infpert' 
+\phipert' 
 +\frac{\partial_x^2 \Lauxpert}{\phib'}
 -\frac{\gamma^2 }{8} \frac{e^{2\phib}\VV_{,\infsub}(\infb' )}{\phib'}\,\infpert  ~.
\nonumber
\eea
Substituting into~\pref{Squadeff} (which is allowed because these variables are non-dynamical), one finds 
\begin{multline}
\label{Stwo}
\SS_2 = \frac{1}{8\pi}\int \biggl( 	
(\infpert')^2 - (\partial_x \infpert)^2 + \slsup^2 (\Lauxpert')^2 - \slsup^2 (\partial_x \Lauxpert)^2
\\
+\,2\slsup\big[\infpert'\Lauxpert'-(\partial_x \infpert)(\partial_x \Lauxpert)\big]
\\
+\,\Bigl[ 2 \Bigl(1-\frac{\gamma^2 }{4}\slsup^2 \Bigr)\infb' +\frac{e^{2\phib}\VV_{,\infsub}(\infb )}{\phib'} \Bigr] \infpert\Lauxpert'
\\
-\frac{\gamma^2 }{2} \(1-\frac{\gamma^2 }{4}\slsup^2 \)(\infb')^2 \infpert^2 
\\
-\frac{1}{2}\[ \gamma^2 \slsup e^{2\phib}\VV_{,\infsub}(\infb )+ e^{2\phib}\VV_{,\infsub\!\infsub}(\infb )\]\infpert^2
\biggr)
\end{multline}
(again up to total derivatives).  
Note that the effective action is independent of $\phipert$.

Having solved the constraints, the resulting action should be expressible in terms of gauge invariant quantities.
Under linearized gauge transformations 
\be
\label{coordtransf}
t\to t+\xi^t(t,x)  ~~, \quad x\to x+\xi^x(t,x) ~,
\ee
the fields transform as
\bea
\label{fieldtransf}
\ntpert &\to& \ntpert + \partial_t{\xi^t} - \partial_x \xi^t +\partial_t \xi^x - \partial_x \xi^x
\nonumber\\
\nxpert &\to& \nxpert - \partial_x\xi^t + \partial_t{\xi^x}
\nonumber\\
\phipert &\to& \phipert + \phib'\xi^t + \partial_x\xi^x
\\
\Lauxpert &\to& \Lauxpert + \Lauxb' \xi^t
\nonumber\\
\infpert &\to& \infpert + \infb' \xi^t
\nonumber
\eea
As in 4d, the quadratic effective action depends only on the analogue of the gauge invariant Mukhanov-Sasaki variable%
~\cite{Sasaki:1986hm,Mukhanov:1988jd}
\be
\label{MS2d}
v = \infpert+\slsup\Lauxpert ~,
\ee
in terms of which one can rewrite~\pref{Stwo} as
\bea
\label{Meffact}
\SS_{2} = \frac{1}{8\pi}\int\biggl( (v')^2-(\partial_x v)^2 + \frac{z''}{z} v^2 \biggr)
\eea
where
\be
z = \slsup ~.
\ee

In fact, the map between perturbations of our 2d theory and the scalar sector of 4d cosmological perturbations can be made quite precise.  A standard parametrization of the scalar metric perturbations in 4d is%
~\cite{Baumann:2009ds,Mukhanov:1990me}
\be
\label{scalarmet}
ds^2_{4d} = a(t)^2\left[-(1+2\Phi)dt^2 + 2B_{,i} dx^i dt + [(1-2\Psi)\delta_{ij}+2E_{,ij}]dx^i dx^j \right] ~.
\ee
Under linearized gauge transformations
\be
\label{coordtransf4d}
t\to t+\xi^t(t,x)  ~~, \quad x^i\to x^i +\partial_i\xi(t,x) ~,
\ee
the 4d scalar modes transform as
\bea
\Phi &\to& \Phi + \partial_t\xi^t + \frac{a'}{a} \xi^t
\nonumber\\
B &\to& B - \xi^t + \xi'
\nonumber\\
E &\to& E + \xi
\\
\Psi &\to& \Psi - \frac{a'}{a} \xi^t  
\nonumber\\
\dX4d &\to&  \dX4d + \hat X' \xi^t  ~.
\nonumber
\eea
One can arrange combinations of the fields $(\Phi,\Psi,E,B,\delta X)$ such that the transformations~\pref{coordtransf4d} cancel, leading to the gauge invariant combinations (the so-called Bardeen potentials)
\bea
\label{invariants4d}
\Phi_B &=& \Phi + \frac1a \left[(B-E')a\right]' 
\nonumber\\
\Psi_B &=& \Psi - \frac{a'}{a}(B-E') 
\\
\inflaton_B &=& \dX4d + \infb'(B-E') ~.
\nonumber
\eea

Comparison to the linearly perturbed 2d metric~\pref{metparam}
\be
ds^2_{2d} = e^{2\phib}\left[ -\left(1+2 (\phipert+\ntpert)\right)dt^2 + 2  \nxpert dx \,dt + 2  \phipert dx^2\right] ~,
\ee
together with its gauge transformation properties~\pref{coordtransf}, suggests the classical identifications
\bea
\label{varmap}
\ntpert &~\leftrightarrow~& \Phi + \Psi - \partial_x^2 E 
\nonumber\\
\nxpert &~ \leftrightarrow~& \partial_x B
\nonumber\\
\phipert &~ \leftrightarrow~& -\Psi + \partial_x^2 E
\\
\Lauxpert &~ \leftrightarrow~& \Psi
\nonumber\\
\infpert &~ \leftrightarrow~& \dX4d  ~.
\nonumber
\eea
%(in making the identification of the gauge transformation properties, one must use the equation of motion $\hat\nabla^2(\phipert+\Lauxpert)=0$).
Including the auxiliary field $\Laux$, the gravitational sector in 2d has as many scalar modes as that of 4d; $\Laux$ provides the necessary fourth scalar mode in this correspondence.  The 2d analogue of the Mukhanov-Sasaki variable~\pref{MS2d} is thus in fact {\it precisely} the same as its 4d analogue
\be
v_{4d} = a\Bigl( \dX4d + \frac{\infb'}{a'/a} \, \Psi \Bigr) = a \Bigl( \inflaton_B + \frac{\infb'}{a'/a} \, \Psi_B \Bigr)  ~,
\ee
apart from a canonical normalization factor of $a=e^\phib$; while similarly the quantity $z$ again differs by a canonical scaling by $a$
\be
z_{4d} = a^2 \infb'/a' ~.
\ee
This different factor of $a$ results in a time-dependent tachyonic mass $z''/z=-2t^{-2}$ for the perturbation $v_{4d}$ in four-dimensional slow-roll, whereas in two dimensions $z''/z=0$ and the variable $v$ is an ordinary massless free field.  This difference in scaling is just what is needed to have a scale-invariant spectrum in both cases.

Since $v$ is a standard, conformally invariant free scalar field in the slow-roll approximation, the physical fluctuation spectrum in 2d is scale invariant, as in higher dimensions.  However the normalization of the 2d fluctuations is of order
\be
|\delta X_\qu| \sim O(1)
\ee
rather than of order $H$ as in 4d (as one might have predicted on the basis of dimensional analysis).  In the quintessence model~\pref{Squint}, the condition that $\delta X_\qu\gg \delta X_\cl$ is simply the slow-roll condition $\beta/\gamma \ll 1$, and so slow-roll eternal inflation predicts that the inflaton is always trying to climb its potential.

For completeness, it is worth displaying the full set of equations of motion in the standard 4d parametrization of the metric.  This exercise extends to the fluctuations the remarkable parallel between Liouville cosmology in 2d and Einstein cosmology in 4d, observed at the level of homogeneous backgrounds in~\cite{DaCunha:2003fm}, see equations~\pref{friedmann}.
Using the identifications~\pref{varmap}, the perturbed Hamiltonian and momentum constraint equations for the quadratic effective action~\pref{Squadeff} read
\bea
%\label{Hamcon4d}
\delta\HH &=& 0 = \frac{4}{\gamma^2} \Bigl[ \phib' \Psi' + (\phib')^2  \Phi - \partial_x^2 \bigl( \Psi - \phib' (B-E') \bigr) \Bigr]
+ \Bigl[ \infb' \dX4d'+\half e^{2\phib} \hat\VV_{,\infsub}\, \dX4d 
- (\infb')^2 \Phi \Bigr]
\nonumber \\
%
%\label{momcon4d}
\label{constraint4d}
\delta\QQ &=& \int^{x}\!\!\! \delta\PP = 0 =  \frac{4}{\gamma^2} \Bigl[\Psi' +\phib'  \Phi   \Bigr] - \infb' \delta X ~.
\eea
These two can be combined into a constraint equation for the Bardeen potential $\Psi_B$ which has precisely the same form as its counterpart in Einstein gravity
\be
\label{PsiBcon}
\partial_x^2 \Psi_B =  \frac{\gamma^2}{4} \( \delta\HH_\matt - \phib' \, {\delta\!\!\; \QQ_\matt} \) ~.
\ee
% This one is just like 4d.

The dynamical equations of motion can also be written in forms closely resembling their 4d counterparts; under the map~\pref{varmap}, the $\phipert$ equation of motion becomes, after using the Hamiltonian and momentum constraints,
\be
\label{phieom4d}
 \Psi'' + \phib' ( \Psi' + \Phi' ) - \bigl( (\phib')^2-\phib'' \bigr)  \Phi  = 
 \frac{\gamma^2}{4} \Bigl[ \infb'\( \dX4d' - \phib' \dX4d \) 
-\frac12 e^{2\phib}\hat\VV_{,\infsub}\, \dX4d \Bigr]
\ee
which is the counterpart of the trace component of the spatial Einstein equations $\delta G_{ij} = \delta T_{ij}$; the traceless part of these equations also has an analogue in the $\Lauxpert$ equation of motion
\be
\label{chieom4d}
0 =  \Phi + \Psi + B'  - E'' 
= \Phi_B + \Psi_B  ~;
\ee
and finally, one has the inflaton equation of motion
\be
\label{Xeom4d}
0 = -\dX4d'' +\partial_x^2 \dX4d - \half e^{2\phib}\hat\VV_{,\infsub\!\infsub}\, \dX4d
+\infb' \( \Phi' + \Psi'  - \partial_x^2 E' + \partial_x^2 B \) - e^{2\phib}\hat\VV_{,\infsub} \, \Phi ~~.~~
%\nonumber
\ee
The equations of motion~\pref{phieom4d}-\pref{Xeom4d} and constraints~\pref{constraint4d} are direct analogues of the equations of linearized cosmological perturbation theory in four dimensions~\cite{Baumann:2009ds,Mukhanov:1990me}.  The expressions above differ slightly because the coefficients of various terms are dimension dependent. 

The correspondence between two- and four-dimensional scalar perturbations means that one has available in two dimensions all the standard gauge choices used in cosmological perturbation theory.  Some standard gauges are
\begin{itemize}
\item
{\it Newtonian gauge}.  Here one chooses $B=E=0$, so that the metric is diagonal.  The Bardeen potentials~\pref{invariants4d} simplify, and the $\chi$ equation of motion sets $\Phi=-\Psi$ in 2d (similarly $\Phi=\Psi$ in 4d if there are no anisotropic stresses), while $\Psi$ is determined by the Gauss-type constraint~\pref{PsiBcon}.  The physical fluctuation is the inflaton $\dX4d$.
\item
{\it Uniform density gauge}.  One chooses a time slicing such that the inflaton is a global clock, $\dX4d=0$; and spatial reparametrizations allow one to set $E=0$ (in 2d, one could alternatively set $\phipert=0$).  This gauge is singular if the classical inflaton velocity $\infb'$ vanishes, because one cannot then adjust the time slicing forward or backward to eliminate $\dX4d$ (\ie\ the gauge slice fails to be transverse).  In this gauge, the constraints can be solved for $\Phi$ and $B$; the physical, fluctuating degree of freedom is $\Psi$, which in four dimensions is the scalar curvature perturbation 
\be
\label{curvepert}
\RR = -\Psi - \frac{\phib'}{{\bar\rho'}} \delta\rho ~.
\ee
The scalar $\RR$ determines the scalar curvature perturbation via $R^{(3)} = (4/a^2)\nabla^2\RR$.  Of course, in 2d there is no spatial curvature, nevertheless $\RR$ is a gauge invariant observable equal to $-\Psi$ in this gauge.
\item
{\it Spatially flat gauge}.  Here one sets $\Psi=E=0$ (\ie\ $\phipert=\Lauxpert=0$ in 2d), so that the spatial scale factor is a global clock, and solves the constraints for the lapse and shift perturbations $\Phi$ and $B$.  This gauge is singular when the expansion rate $\phib'$ vanishes, such as at the \dS bounce.  The dynamical variable is the inflaton fluctuation $\dX4d$, and the observable $\RR$ in~\pref{curvepert} is determined by the matter density fluctuations rather than directly in terms of the metric.  Discussions of slow-roll eternal inflation usually take place in this gauge.
\item
{\it Synchronous and conformal gauges}. Gauge choices where one fixes the Lagrange multipliers for the gauge constraints, such as synchronous gauge
\be
\Phi=B=0
\ee
or conformal gauge
\be
\ntpert = \Phi+\Psi-\partial_x^2E = 0~~,\quad \nxpert = \partial_x B = 0  ~,
\ee
leave additional fluctuating fields to be quantized, and the constraints that eliminate the unphysical ones are only imposed weakly on the space of states, after quantization.  This leads to a potential difficulty -- there are negative metric fluctuations, and so instabilities are a concern.  In string theory, one typically deals with this issue through analytic continuation of the fields to/from some regime where the path integral is convergent. 
\end{itemize}
This last point is underscored by the fact that $\phipert$ in 2d, and $E$ in 4d, are absent from the gauge-invariant effective action~\pref{Meffact}.  We will see below that Liouville perturbation theory suffers large infrared divergences; this does not mean that \dS timelike Liouville theory is doomed, but it does mean that a more sophisticated approach is called for.

Physical gauges such as spatially flat or uniform density gauge are convenient, in that the only fluctuating quantities are physical degrees of freedom.  The quadratic effective action~\pref{Meffact} is written directly in terms of the gauge invariant quantity $v$ for which we can choose as representative either the spatial metric perturbation $\Psi$ or the density perturbation $\delta\rho$ (which is basically $\dX4d$).  These gauges have a sensible perturbation theory, since part of the gauge choice is $E=0$ in 4d, or $\phipert=0$ in 2d.  However, the expansion of the effective action in these gauges is cumbersome; it may be worth braving the subtleties and pitfalls of conformal gauge, if one can make sense of \dS timelike Liouville theory -- the action is quite simple, even if its quantization is subtle.  We will explore some aspects of conformal gauge quantization after we have related our two basic models~\pref{STL} and~\pref{Squint}.

%%%%%%%%%%%%%%%%%%%%%%%%%%%%%%%%%%%%%%%%%%
%%%%%%%%%%%%%%%%%%%%%%%%%%%%%%%%%%%%%%%%%%

\section{Relating quintessence to pure \dS gravity}
\label{sec:ModelEquivalence}

It turns out that the two models discussed above -- the \dS timelike Liouville theory~\pref{STL} and the Liouville-quintessence model~\pref{Squint} -- are related by a field redefinition.  In this section, we exhibit the field redefinition and point out that the Liouville quintessence appears to provide a counterexample to the phenomenon of slow-roll eternal inflation.  A generalization to a cosh potential for the inflaton provides a further counterexample.  We then proceed to discuss a variety of potential objections to this line of reasoning.  We end with a brief exploration of four-dimensional analogues of the Liouville-quintessence model.

%%%%%%%%%%%%%%%%%%%%%%%%%%%%%%%%%%%%%%%%%%
%%%%%%%%%%%%%%%%%%%%%%%%%%%%%%%%%%%%%%%%%%

\subsection{Field redefinitions}

The quintessence model \pref{Squint} with fields $\tilde\phi$, $\tilde\inflaton$ is related to Liouville theory~\pref{STL} with fields $\phi$, $\inflaton$ by a `boost' field redefinition%
\footnote{Our conventions for which fields carry a tilde are henceforth reversed relative to the introduction.}
\be
\label{fieldredefn}
\begin{pmatrix}2\tilde{\phi}/\gamma\\ \tilde{X}\end{pmatrix}=M(\boost)\begin{pmatrix} {2\phi/\gamma}\\   X\end{pmatrix},\qquad{\;}
M(\boost)=\begin{pmatrix}\cosh\boost & ~\sinh\boost\\ \sinh\boost & ~\cosh\boost\end{pmatrix} ~,
\ee
which leaves the kinetic term and the path integral measure invariant, and relates the parameters in the respective actions via
\bea\label{newparams}
\left(\begin{matrix} 2\tilde Q_\phican/\gamma \\ \tilde Q_\infsub \end{matrix}\right) = 
M(\boost) \left(\begin{matrix} 2Q_\phican/\gamma \\ Q_\infsub \end{matrix}\right)  ~~,\quad
\left(\begin{matrix} \tilde\alpha \\ \tilde\beta/\gamma \end{matrix}\right) =
M(\boost) \left(\begin{matrix} 1 \\ 0 \end{matrix}\right) 
\eea
(in the modified Liouville, theory, one must also shift the auxiliary field $\chi$ to maintain the form of the action).
Thus the potentials in the two frames are related by
\be
\label{twoVs}
e^{2\phi} = e^{2\tilde\alpha \tilde\phican - \tilde\beta \tilde \inflaton}
\ee
with $\tilde\alpha=\cosh\lambda$, $\tilde\beta=\gamma\sinh\lambda$.
For $\sinh(\boost)\ll1$, the slow-roll conditions~\pref{slowroll} are satisfied.  With this choice, we can immediately write down the solution of the model at linearized order using the results of the previous section.

The field redefinition~\pref{fieldredefn} relates the background solution for quintessence to that of the \dS solution of Liouville theory~\pref{dsmetrics}, plus an additional free field.  Let us choose non-compact spatial sections; then the quintessence solution is
\bea
\label{boostfields}
\frac2\gamma \tilde\phi &=& -\frac{\cosh\boost}\gamma \log[\Lambda t^2] + \sinh\boost ~\infb \\
\tilde\inflaton &=&  -\frac{\sinh\boost}\gamma \log[\Lambda t^2] + \cosh\boost ~\infb ~.
\eea
This classical solution has the property that $\tilde\inflaton\to +\infty$, \ie\ small potential, as one evolves to large volume $\tilde\phi\to +\infty$, as one would expect.  If, on the other hand, slow-roll eternal inflation is in operation, one expects the dominant measure for $\tilde\inflaton$ at large volume $\tilde\phi\to\infty$ to be concentrated on configurations with $\tilde\inflaton\to -\infty$.   The field redefinition~\pref{fieldredefn} then implies the behavior 
\be
X =  \cosh\boost\, \,\tilde X - \frac2\gamma \sinh\boost\, \,\tilde\phi \to -\infty
\ee
of the free matter field $\inflaton$ in Liouville gravity.
For $\tilde\inflaton$ to climb its potential via slow-roll eternal inflation implies, in the other frame, the presence of a mysterious drift force that pushes the free field $\inflaton$ to large negative values $\inflaton\to -\infty$.  It is not clear how such a drift force could arise, since in conformal gauge the free field $\inflaton$ is completely decoupled from the gravitational sector.  

Note that nowhere in this argument does one need to invoke the small fluctuation expansion elaborated above in section~\ref{sec:fluctuations}.  Thus, on the one hand, the class of two-dimensional models being considered has exactly the same perturbative content and structure of scalar fluctuations as inflationary models in higher dimensions, and for an appropriate parameter regime satisfies the criteria for slow-roll eternal inflation; on the other hand, it is related by an exact field redefinition to free field theory in \dS space.  In the latter description, the sorts of fluctuations predicted by slow-roll eternal inflation do not seem to occur.  It is hard to see how slow-roll eternal inflation could arise in this model.  

%%%%%%%%%%%%%%%%%%%%%%%%%%%%%%%%%%%%%%%%%%
%%%%%%%%%%%%%%%%%%%%%%%%%%%%%%%%%%%%%%%%%%

\subsection{More general models}
\label{sec:generalizations}

The quintessence model is free field theory in disguise, therefore it has many hidden symmetries.  One might object that these symmetries secretly suppress eternal inflation in the 2d quintessence model, but since they are not generic, the mechanism of slow-roll eternal inflation might still be valid in general.  However, already this argument should give one pause, since it suggests that the standard argument involving gravitational back-reaction of gravity on decohered matter fluctuations would need to be refined in light of this subtlety, since it would seem to apply as well to the quintessence model.

We believe that such hidden symmetries are not the reason for the apparent absence of slow-roll eternal inflation.  Instead of a quintessence potential as in~\pref{Squint}, consider a potential with several exponentials.   There will then be no field redefinition that exhibits a large class of hidden symmetries, and yet field redefinitions lead to apparent contradictions with the slow-roll eternal inflation paradigm.

For example, let the potential for the inflaton be
\be
\label{coshpotl}
\tilde\VV = \mu^2 e^{2\tilde\alpha\tilde\phican} \cosh(\tilde\beta\tilde\inflaton)
\ee
with $\tilde Q_\infsub=0$ so that both exponentials in the cosh have the same scale dimension.
Again, for $\tilde\beta$ sufficiently small, according to the standard criteria this potential exhibits slow-roll eternal inflation where the inflaton $\tilde\inflaton$ is driven to large positive values, or large negative values, depending on whether the initial condition is placed at positive or negative $\tilde\inflaton$; on the other hand, classical evolution always predicts that the inflaton $\tilde\inflaton$ is driven to zero at large scale factor $\tilde\phican$. 

We can now apply boost field redefinitions that convert either of the exponentials into a pure cosmological constant term; the transformation~\pref{fieldredefn} that leads to~\pref{twoVs} for instance relates the original potential~\pref{coshpotl} to
\be
\label{CCandExp}
\VV = \frac{\mu^2}2 \left(e^{2\phi} + e^{2\alpha\phican} e^{ \beta\inflaton} \right) 
~~,\qquad
\left(\begin{matrix}\alpha \\\beta/\gamma \end{matrix}\right) =
M(\boost) \left(\begin{matrix} \tilde\alpha \\ \tilde\beta/\gamma \end{matrix}\right) ~.
\nonumber
\ee
In the redefined frame, slow-roll eternal inflation predicts that $\inflaton$ is always driven to $+\infty$ at large scale factor $\phi$, independent of the initial value of $\inflaton$, whereas classical evolution predicts that $\inflaton$ is driven to $-\infty$.  If we had boosted in the opposite direction, we would have predicted that there is only quantum jumping in the direction of minus infinity, whereas classical evolution drives the redefined field to plus infinity.

Translating back to the evolution in the cosh frame $\tilde\phi$, $\tilde\inflaton$, slow-roll eternal inflation in $\inflaton$ predicts the dynamics for $\tilde\inflaton$
\be
\tilde\inflaton =  \cosh\boost\, \, X + \frac2\gamma \sinh\boost\, \, \phi \to +\infty ~~,
\ee
independent of the initial value of $\tilde\inflaton$.   If we had performed the boost in the opposite direction, we would have concluded that the field $\tilde\inflaton$ is always driven to minus infinity, independent of its initial value.  

Thus, the issue is not whether one has a model that is secretly free field theory; rather it's that the logic of slow-roll eternal inflation is internally inconsistent -- when one applies that logic to the same theory viewed in different coordinate frames in field space, one predicts different outcomes, not remotely compatible with one another. 

%%%%%%%%%%%%%%%%%%%%%%%%%%%%%%%%%%%%%%%%%%
%%%%%%%%%%%%%%%%%%%%%%%%%%%%%%%%%%%%%%%%%%

\subsection{Possible objections}
\label{sec:objections}

The preceding result flies in the face of conventional wisdom.  Therefore, it behooves us to consider whether there are any subtleties that might prevent the above quintessence model from being a counterexample to the logic of slow-roll eternal inflation.  Let us list here a few:

\vskip 0.3cm
\noindent
{\it {The theory might not exist}.}  

It is not clear that the action~\pref{STL} describes a conformal field theory.  The exponential potential solves the conformal invariance condition at semiclassical order, but it is not clear that this remains true at large $\phi$ where the exponential becomes large; yet this is precisely this late-time \dS regime of large scale factor that we wish to consider.  In the semiclassical spacelike Liouville theory ($c_L$ large and negative, and the AdS sign of the cosmological constant), the exponential is self-consistent for small Liouville energy $\varepsilon$ since the dynamics never explores the regime of large values of the potential.  The weak-coupling perturbation expansion is fully self-consistent.  The status of timelike Liouville theory, with the \dS sign of the cosmological constant, is much less clear.

There have been a few investigations of the structure of the beta functions in the presence of potential interactions (a `tachyon background' in the language of string theory). %~\cite{Banks:1991sg,Tseytlin:2000mt,Headrick:2008sa}.
In two dimensions, the generalized beta functions that determine the conditions for scale invariance are thought to be derivable from a variational principle on the space of couplings (this is the target space effective action in the string theory interpretation of the 2d dynamics).
There are claims~\cite{Banks:1991sg} that, due to field redefinition ambiguities, one could take the form of the field space effective potential for the 2d coupling $\VV(X)$%
\footnote{Note that there are two uses of the notion of `potential' here which need to be distinguished.  The 2d potential $\VV(X)$ is a coupling in the 2d field theory describing dynamics in 2d `spacetime'.  There is also an effective potential in the space of couplings $\VV$, $\GG$, $\Phi$ \etc, which governs which forms of $\VV(X)$ lead to consistent quantum theories of 2d gravity; it is the properties of this effective potential in the space of couplings that is the subject of the present discussion.} 
in this field space effective action to be given exactly by $-\VV^2$, which would seem to indicate that the growth of $\VV$ never subsides; on the other hand,~\cite{Tseytlin:2000mt} argued for a modification to $-\VV^2 e^{-\VV}$, which would seem to lead to an endpoint to `tachyon condensation', since the potential for $\VV$ has a minimum.  A more recent analysis~\cite{Headrick:2008sa} showed that there is no target space effective action involving the couplings $\VV$, $\GG$, and $\Phi$ of~\pref{sigmodel} satisfying certain expected properties -- that when the 2d action is a sum of decoupled field theories, the beta functions factorize in an appropriate way.  Tachyon dynamics was also studied in~\cite{Freedman:2005wx,Suyama:2005wd,Hellerman:2006nx,Swanson:2008dt,Adams:2009yb}.  The situation seems at the moment rather murky.  

For our purposes, what is needed is that there is no endpoint to `tachyon condensation' in the coupling space -- that the exponential growth of the potential $\VV$ persists for an arbitrarily long time, so that the 2d dynamics is some approximation of \dS geometry all the way out to arbitrarily large $\phi$, and thus arbitrarily large spatial volume.%
\footnote{Of course, if there were an endpoint to tachyon condensation, that would also be interesting, as it would be an example of a mechanism for dynamical relaxation of the cosmological constant through quantum effects.}  
Actually, this property may not be completely necessary -- it should be sufficient that there is some sufficiently long epoch in which the 2d classical dynamics looks like \dS expansion, and there are many e-foldings of that expansion during which the criteria for slow-roll eternal inflation are met, while at the same time we can find a field redefinition that orients the gradient of $\VV$ along the $\phi$ direction (this would be the approximately decoupled `Liouville plus free field' frame).

Note that the late time properties of local (sub-Hubble scale) observables in this class of models is intimately bound up with the ultraviolet behavior of the Liouville field theory at large positive $\phican$, since fixed proper distance becomes ever smaller coordinate separation as the scale factor continues to grow.  Thus it would be helpful to understand the nature of closed string tachyon condensation in string theory well above the critical dimension $c_\matt\gg 25$ (or in the modified Liouville theory, which is critical string theory with a large light-like dilaton).

\vskip 0.3cm
\noindent
{\it {Timelike Liouville theory might exist, but not have the requisite properties}.}  

The timelike Liouville theory that appears in conformal gauge has been the subject of a number of investigations~\cite{Schomerus:2003vv,Zamolodchikov:2005fy,Kostov:2005kk,Kostov:2005av,McElgin:2007ak,Harlow:2011ny}, which have sought to adapt to the timelike regime ($c_L\le 1$) the conformal bootstrap technology that solves the spacelike ($c_L \ge 25$) Liouville theory.  The bootstrap considers the properties of correlation functions involving insertions of a class of degenerate operators (operators having a null vector in its tower of descendants under the action of the conformal algebra).  Conformal Ward identities then lead to constraints on the correlators.  Additional analytic properties of the correlators, such as crossing symmetry and factorization, together with the conformal Ward identities and the assumption of a unique operator of each conformal highest weight, lead to a set of discrete functional identities on correlators.  For $c_L\ge 25$, these relations are sufficient to uniquely specify the correlation functions~\cite{Dorn:1994xn,Zamolodchikov:1995aa}.  For $c_L\le 1$, the corresponding exercise leads to two candidate solutions for the correlation functions~\cite{Zamolodchikov:2005fy,Schomerus:2003vv,Kostov:2005kk,Kostov:2005av,McElgin:2007ak,Harlow:2011ny}, neither of which satisfies all the expected properties of a conformal field theory such as vanishing of the two-point function for two conformal fields of different scale dimension.
Thus, at the moment there are unresolved issues with the conformal bootstrap.  Even if these are resolved, one will then need to understand whether these largely Euclidean techniques are applicable to the intrinsically Lorentz signature issues being addressed here.

Nevertheless, the semiclassical expansion around the \dS background seems no {\it less} consistent than in higher dimensions, and has the additional advantage of perturbative renormalizability.  We have seen that there is a precise map between the auxiliary field Liouville-quintessence model in two dimensions, and the scalar sector of cosmological perturbation theory in four dimensions.  In both cases there is a scale-invariant spectrum of linearized perturbations, and the small fluctuation expansion is quite similar in structure.  At the perturbative level, the theory seems fully consistent, provided that there is a scheme to renormalize UV divergences, and a method to regulate and resum IR singularities.

\vskip 0.3cm
\noindent
{\it There might not be an interpretation in terms of a single universe.} 

Since 2d gravity can also be interpreted as string theory, there is the issue of topology change in the 2d geometry -- it would be difficult to give  a single-universe interpretation to the dynamics if there is a significant probability of a catastrophic topology-changing event occurring in a given spatial domain.  The coupling $\Phi(X)\R$ in the 2d action can be thought of as governing the amplitude for topology-changing processes for string worldsheets, which here represent an ensemble of 2d cosmologies.  In timelike Liouville theory, the coupling $\hat\R \phi$ enhances the likelihood of topology change at large negative $\phi$ where the exponential Liouville potential is vanishingly small, and exponentially suppresses topology change in the regime of large positive $\phi$ which governs the late-time \dS dynamics.  Thus, while the early history of the 2d cosmology at small spatial volume may be rife with topology-changing processes, at late time and large spatial volume, these processes are highly suppressed.  If one adopts a Hartle-Hawking ansatz for the generation of the expanding universe through a tunneling process, the region of strong topological fluctuation is in an exponentially suppressed region of field space.

On the other hand, consider the rate of string pair production in timelike Liouville theory.  This rate was estimated in~\cite{Strominger:2003fn}, where it was shown that string pair production is exponentially suppressed as $\Gamma \sim e^{-2\pi \varepsilon/\gamma}$ for a string (\ie\ 2d universe) with energy (Liouville momentum) $\varepsilon$.  However it is exponentially enhanced by the density of states $\rho \sim e^{4\pi \varepsilon/\gamma}$,%
\footnote{This is the density of states when the spacelike fields have no (or very small) conformal improvement terms in their stress tensor; large conformal improvement terms of these fields will reduce the density of states.  For instance, in the modified Liouville theory with light-like dilaton, the density of states only grows as $e^{4\pi\varepsilon}$ and for small enough $\gamma$ the pair production rate is finite.}
and so the total rate of production $\rho \Gamma$ diverges.  If we are forced to think about an ensemble of universes, according to this analysis that ensemble is dominated by the proliferation of highly excited states in the Hilbert space of a single universe.%
\footnote{The back-reaction of this proliferation of excited strings was argued in~\cite{Aharony:2006ra,Frey:2008ke} not to cause a significant perturbation of the background at late times.}

\vskip 0.3cm
\noindent
{\it Conformal gauge might be pathological.}  

The disparity between the predictions of slow-roll eternal inflation and conventional semiclassical dynamics is particularly stark in conformal gauge for the quintessence model~\pref{Squint}, where in the redefined frame one has a decoupled Liouville field and free field.  The conclusion might be on somewhat shakier ground if conformal gauge were somehow pathological; and indeed, we will see below that the small fluctuation expansion is problematic.  Outside of conformal gauge, there is a coupling between the gravitational sector and the matter sector through the curvature term $\hat\R \inflaton$, which now contains dynamical fields.  One could then worry that fluctuations of $\hat\R$ provide some sort of drift force that indeed pushes the inflaton $\inflaton$ in a preferred direction through an effective potential $\langle\hat \R\rangle \inflaton$.  
However, one can choose to start with a conformal improvement in the quintessence frame which is cancelled by the field redefinition, leaving $\inflaton$ as an ordinary free field with no curvature coupling, or one can choose either sign of this coupling through an appropriate value in the quintessence frame.  It seems unlikely that the effects of the improvement term alter the conclusion.  Furthermore, the more general example of the cosh potential doesn't really need conformal gauge to exhibit an internal inconsistency in the logic of eternal inflation.

%%%%%%%%%%%%%%%%%%%%%%%%%%%%%%%%%%%%%%%%%%%%%%%
%%%%%%%%%%%%%%%%%%%%%%%%%%%%%%%%%%%%%%%%%%%%%%%

\subsection{Four-dimensional models}

The two-dimensional models parallel the structure of four-dimensional cosmology quite closely.  However, one might wish to study directly a four-dimensional cosmological model exhibiting the same structure as the 2d quintessence model explored above.  Are there four-dimensional quintessence models similarly related to free field theory?

Consider the class of 4d metrics
\be
ds^2 = e^{2\phi} \ghat_{ab} = e^{2\phi} \[ (-N^2 + N_i N^i)dt^2 + 2N_i dt dx^i + \ghat_{ij} dx^i dx^j \]
\ee
where spatial indices are raised and lowered with the unit determinant metric $\ghat_{ij}$.
The 4d Einstein action coupled to a scalar inflaton $X$ is
\be
\label{Sein}
\SS_{4d} = \frac{\mpl^2}{2 } \int \sqrt{- \ghat} \,e^{2\phi}\[ \hat R + 6 (\hat\nabla\phi)^2 - 6(\hat \nabla X)^2 - e^{2\phi}\, \VV(X) \] ~,
\ee
where to simplify further developments we have chosen a non-canonical normalization for $X$.  The analogue of (modified) timelike Liouville theory is Einstein gravity with a cosmological constant $\VV(X)=6\Lambda$, and we similarly couple it to a free scalar field.

The gradient terms in square brackets again are invariant under a boost transformation
\be
\label{redef4d}
\begin{pmatrix}{\tilde\phi} \\ \tilde{X} \end{pmatrix}=M(\boost)\begin{pmatrix} {\phi}\\   X\end{pmatrix}
\ee
under which the action transforms into
\be
\label{Squint4d}
\SS_{4d} = \frac{\mpl^2}{2} \int \sqrt{- \ghat} \,e^{2(\atil\phitil-\btil \Xtil)}\[ \hat R + 6 (\hat\nabla\phitil)^2 - 6(\hat \nabla \Xtil)^2 - e^{2 (\atil\phitil-\btil \Xtil)}\, \VV(\atil \Xtil - \btil\phitil) \] ~,
\ee
with $\atil=\cosh\lambda$, $\btil=\sinh\lambda$.    This action is the analogue of the Liouville-quintessence model~\pref{Squint}.  Unfortunately, it cannot be written in Einstein frame -- that would get us back to~\pref{Sein}.  However, this canonical normalization is not necessary for investigating the logic of slow-roll eternal inflation.  In the slow-roll approximation in a Jordan frame such as~\pref{Squint4d}, both the effective gravitational coupling and the Hubble scale are evolving slowly.  The standard argument, that quantum fluctuations trump classical displacement of the inflaton, is not affected by this slow evolution of the Newton constant, any more than it is affected by the slow classical evolution of the Hubble scale; both are slow-roll suppressed effects, and the claim is that quantum fluctuations of the inflaton far outweigh them in the eternal inflation regime.  We have
\be
\mpleff^2 = \mpl^2 e^{2 [(\atil-1)\phitil_0-\btil \Xtil_0]} ~~,\quad   \mpleff^2H^2_\eff \approx \Lambda  e^{4 [(\atil-1)\phitil_0-\btil \Xtil_0)] }
\ee
where $\phitil_0(t)$, $\Xtil_0(t)$ constitute the slowly evolving background values, locally constant over a Hubble volume.  
To leading order in the slow-roll approximation, the equation of motion of $\Xtil$ is
\be
2 H_\eff \, {\dot\Xtil}_0 \approx \beta\Lambda e^{2[(\alpha-1)\phitil_0-\beta\Xtil_0]} ~,
\ee
and thus the slow-roll eternal inflation criterion~\pref{SREIcriterion} is then met when
\be
\frac{\mpleff^2\dot \Xtil^2}{H_\eff^{4}} \approx \btil^2\frac{\mpl^2}{\Lambda} \ll 1~.
\ee
Again, the standard paradigm predicts slow-roll eternal inflation where clearly it does not happen.

%%%%%%%%%%%%%%%%%%%%%%%%%%%%%%%%%%%%%%%%%%
%%%%%%%%%%%%%%%%%%%%%%%%%%%%%%%%%%%%%%%%%%

\section{Semiclassical quantization of timelike Liouville}
\label{sec:semiclassical}

In this section, we collect some results on the quantization of timelike Liouville theory in conformal gauge.  A sensible quantization would help allay some of the potential concerns raised above; and provide us with a theory of quantum two-dimensional \dS spacetime.

\subsection{Perturbation theory is insufficient}
\label{sec:nonlinearity}

At small coupling $\gamma$, the semiclassical approximation to the Liouville dynamics should be accurate, however perturbation theory is not.  
Expanding the action~\pref{STL} around a classical solution
\be
\phi = \phib + \varphi ~,
\ee
we choose the background \dS solution
\be
\phib = -\frac12 \log\[\Lambda t^2 \] ~.
\ee
The linearized equation of motion for fluctuations $\varphi$ about this background is
\be
\label{lineareom}
\hat\nabla^2\varphi + 2\Lambda e^{2\phib} \,\varphi = 0 ~.
\ee
The solutions (in momentum space for the spatial coordinate $x$) are the same spherical Bessel functions that appear in the analysis of inflaton dynamics in four dimensions:
\be\label{parpert}
\varphi_k = \half \sqrt{\frac{\pi}{k}} \sqrt{-kt} \,H^{(1)}_{3/2}(-kt) = \frac{1}{\sqrt{2k}}\(1- \frac{i}{kt}\) e^{-ikt}
\ee
and its Hermitian conjugate.  These solutions describe modes which oscillate as positive frequency plane waves in the far past, freeze out at horizon crossing, and then grow as $1/t$ at late times.  The field then has a mode expansion
\be
\varphi(t,x) = \int\frac{dk}{{2\pi}} \(\hat a_k\varphi_k(t)e^{ikx} + \hat a_k^\dagger \varphi^*_k(t) e^{-ikx} \) ~.
\ee

The equal-time position space two-point function is a measure of the quantum fluctuations of the metric.  It has the form
\be
\bra{0} \varphi(t,x) \varphi(t,x') \ket{0} = 
\int \frac{dk}{{2\pi}} \( \frac{(kt)^2+1}{2k(kt)^2} \) e^{ik( x-x')} ~,
\ee
and exhibits a strong infrared divergence at late times.  In particular the fluctuation spectrum of superhorizon modes of the scale factor $\phican$ is not scale invariant.
This fact does not imply a breakdown of this property for physical fluctuations; we have seen above that the gauge-invariant fluctuation spectrum of the geometry {\it is} that of a canonical free scalar field $v$ (with positive, not negative kinetic energy).  Rather, the point is that the fluctuations of $\phican$ are not gauge invariant; moreover, we will see below that their divergences are at worst logarithmic, not power law -- the linearized approximation breaks down well before the power law growth of the linearized solution takes over.  

In gauges such as conformal gauge or synchronous gauge, one first quantizes all the modes of the metric apart from $g_{0a}$ or $N^a$, including the scale factor mode $\phican$ whose kinetic energy has the ``wrong'' sign; then one imposes the positive (negative) frequency components of the constraints on the ``in'' (``out'') states.  However, one will have to solve that quantization problem at the fully nonlinear level, and then impose the gauge constraints on the set of solutions.

Note that, even if the field modes are placed in the Bunch-Davies vacuum at early times $t\to -\infty$, this does not imply that they are in a physical state at late times, due to mode amplification in the time-dependent background.  A similar issue occurs in Hawking radiation along a macroscopic string captured by a black hole~\cite{Lawrence:1993sg}; if the modes are quantized in conformal gauge and placed in the Bunch-Davies vacuum, then there will be Hawking radiation of longitudinal modes and Faddeev-Popov ghost modes as well as the transverse modes of the string.  Without solving the black hole evaporation problem in its entirety, one should at least impose the physical state constraints on the outgoing radiation on the string.  Alternatively, one can work in a physical gauge, and quantize only the transverse modes of the string.  Similarly, when working in conformal gauge for Liouville cosmologies, one should impose the physical state constraints on the late time state; this will tie fluctuations of the conformal mode $\phican$ to those of the matter fields.

The large late-time fluctuations of $\phican$ are not specific to two dimensions.  Consider the linearized fluctuations $\phi = \phib + \varphi$ of the scale factor in the class of 4d metrics of the form $e^\phi \hat g_{ab}$.  They obey a wave equation whose mode solutions are $\varphi_k \propto (-kt)^{3/2}H^{(1)}_{5/2}(-kt)$, and so again blow up as $-1/t$ at late times $t\to 0^-$.  If we are not working in a physical gauge, then to tame these strong, unphysical IR fluctuations, we should understand them at the nonlinear level; and we should impose the constraints at late times.

Liouville perturbations can be understood at the nonlinear level using exact classical solutions.  Any solution of the classical equation of motion~\pref{Leom} can be expressed locally as
\be
\label{Lexact}
 e^{2\phi}=-\frac{4}{\Lambda} \frac{\partial_+{A}\,\partial_-{B}}{(A-B)^2}~,
\ee
where $A(x^+)$ and $B(x^-)$ are arbitrary functions of the left and right moving coordinates $x^\pm$. In other words, the metric $e^{2\phi}\eta_{ab}$ is related to the canonical \dS metric $ds^2=\frac{1}{\Lambda}(-dt^2+dx^2)/t^2$ by a coordinate transformation $x^+\to A(x^+)$, $x^-\to B(x^-)$.
For instance for the homogeneous hyperbolic solution in~\pref{dsmetrics} has
\be
A=e^{\varepsilon x^+} ~~,\qquad B=e^{-\varepsilon x^-} ~~.
\ee
The parabolic and elliptic cases may be obtained from this by taking $\varepsilon\to 0$ or continuing $\varepsilon \to i\varepsilon$.

Consider a plane wave perturbation on the parabolic background
\be
A =  x^{+}+\pertsize e^{-ikx^{+}} ~~,\qquad B=x^- ~~;
\ee
it generates a perturbed Liouville field
\be
\label{fullparpert}
e^{2(\phib+\phichange)}=\frac{1}{\Lambda}\frac{1-\pertsize ike^{-ikx^{+}}}{\(t+\frac{\pertsize}{2}e^{-ikx^{+}}\)^{2}}~~,\qquad
e^{2\phichange} = \frac{1-\pertsize ike^{-ikx^{+}}}{\(1+\frac{\pertsize}{2}\frac{e^{-ikx^{+}}}{t}\)^{2}} ~.
\ee
The perturbation $\phichange$ doesn't satisfy the Liouville equation, but the fact that $\phib$ and $\phib+\phichange$ are solutions implies
\be
\label{dphinonlinear}
\hat\nabla^2(\phichange) +\Lambda e^{2\phib}\(e^{2\phichange}-1\) = \hat\nabla^2(\phichange) +\Lambda e^{2\phib}\(2\phichange+\half(2\phichange)^{2}+\cdots\) = 0 ~.
\ee
The first term in brackets gives the linearized equation of motion~\eqref{lineareom}.  Substituting the form \eqref{fullparpert} of $e^{2\phichange}$ and expanding in $\pertsize$,
\be
1+2\phichange+O(\phichange^{2})=1-\pertsize ike^{-ikx^{+}}\(1-\frac{i}{kt}\)+O(\pertsize^{2}) ~.
\ee
The linear approximation matches $\varphi$ \eqref{parpert}, but at late times the higher order terms in the expansion become important before the power law singularity takes over.  The actual singularities in the Liouville field perturbations are at worst logarithmic (and complex, since we have chosen a complexified perturbation), as~\pref{fullparpert} shows.

The logarithmic singularities of $\phichange$ represent a shift in the coordinate time location of conformal infinity. To see this, consider a general perturbation of the functions $A,B$ which locally describe the Liouville field:
\be
A\to e^{\varepsilon x^+}\(1+\Achange(x^{+})\),\qquad{\,}B\to e^{-\varepsilon x^-}\(1+\Bchange(x^{-})\)~.
\ee
The perturbed Liouville field is
\be\label{pertphi}
e^{2(\phib+\phichange)}=\frac{\varepsilon^{2}}{\Lambda}\frac{\bigl(1+\Achange+(\Achange)'/\varepsilon\bigr)\bigl(1+\Bchange-(\Bchange)'/\varepsilon\bigr)}{\(\sinh\varepsilon t+\half e^{\varepsilon t}\Achange -\half e^{-\varepsilon t}\Bchange\)^{2}}~.
\ee
The singularity in $\phican$ no longer occurs at $t=0$, in general. Instead, each spatial point $x$ reaches a singularity at a time $\tsing{x}$ which solves
\be
\label{bdyloc}
1+\Achange\bigl(\tsing{x}+x\bigr)-e^{-2\varepsilon \tsing{x}}\[1+\Bchange\bigl(\tsing{x}-x\bigr)\]=0~.
\ee
Assuming the perturbations $\Achange, \Bchange$ are bounded, $\tsing{x}$ always exists. 
The singularity of the metric is always of the form $1/(t-\tsing{x})^{2}$; thus we expect perturbations of the Liouville field to grow logarithmically at late times, as the parameter time location of the conformal boundary is shifted by the perturbation.

%%%%%%%%%%%%%%%%%%%%%%%%%%%%%%%%%%%%%%%%%%
%%%%%%%%%%%%%%%%%%%%%%%%%%%%%%%%%%%%%%%%%%

\subsection{The WKB limit}

The minisuperspace approximation to Liouville theory provides some intuition about the form of the wavefunctional as a function of the Liouville field $\phican$.  The zero mode truncation of the Liouville Hamiltonian
\be
H = \frac12\partial_\phican^2 + \Lambda e^{2\phican}
\ee
governs quantum mechanics in an upside down exponential potential, whose energy eigenfunctions are Hankel functions; these have an integral representation
\be
\label{Hankelintrep}
e^{-\varepsilon\pi /2}H_{i \varepsilon}^{(1)}(e^\phican) = \frac{1}{\pi i} \int_{-\infty}^{\infty} e^{-ie^\phican\cosh \psi}\,e^{ - i\varepsilon \psi} d\psi ~.
\ee
The large $\phi$ asymptotics is readily determined by the WKB or saddle point approximation to this integral to be
\be
\label{Hasymp}
e^{-\varepsilon\pi/2}H_{i \varepsilon}^{(1)}(z) \sim \sqrt{\frac{2}{\pi z}}\, e^{iz - i\pi/4} ~.
\ee
One expects a similar structure in the full theory -- the WKB approximation should be accurate at large scale factor $e^\phican$, and also for small coupling $\gamma$.  Indeed there is a remarkable, explicit WKB wavefunctional in the modified Liouville theory~\pref{Lcov} (with $Q=1$).  The Hamiltonian density of this theory is
\be
\label{HmodL}
\HH = \frac{\gamma^2}{4}\Bigl( -\pi_\phican^2 + 2 \pi_\phican\pi_\Laux \Bigr) + \frac{1}{\gamma^2} \Bigl( (\partial_x\Laux)^2 + 2 \partial_x\phican\partial_x\Laux - 2\partial_x^2\Laux + \Lambda e^{2\phican} \Bigr) ~.
\ee
In the WKB approximation, the momenta $\pi_\phican$, $\pi_\Laux$ are given by the functional derivatives $-i\delta\SS_\cl / \delta\phican$, $-i\delta\SS_\cl / \delta\Laux$ of the classical action.  One can readily verify that the ansatz
\bea
\SS_\cl[\phican,\Laux] &=& \frac{2}{\gamma^2}\int dx \( \int^\phican d\varphi \sqrt{\Lambda\, e^{2\varphi} + ( \partial_x\Laux)^2} \) 
\nonumber\\
&=& \frac{2}{\gamma^2} \int dx \biggl( \sqrt{\Lambda\, e^{2\phican} + (\partial_x\Laux)^2} 
- \partial_x\Laux \, {\rm arctanh} \! \biggl[ \frac{ \sqrt{\Lambda\, e^{2\phican} + (\partial_x\Laux)^2} } { \partial_x\Laux} \biggr] \biggr) 
\nonumber
\eea
solves the Hamilton-Jacobi equation of the model for zero energy.
Thus the WKB wavefunctional for the gravity sector at $\varepsilon=0$ is
\be
\Psi_{\varepsilon=0}[\phican,\Laux] \sim  {\exp\Bigl[i\SS_\cl[\phican,\Laux] + \SS_1+ O(\gamma^2)\Bigr]} ~~,
%{ \Bigl({\int} dx  \sqrt{\Lambda\, e^{2\varphi} + (\partial_x\Laux)^2} \Bigr)^{1/2} } 
\ee
where the fluctuation determinant correction $\SS_1$ has the large volume behavior
\be
\SS_1 = -\frac14 \int dx\, \mu \log\Bigl[\Lambda e^{2\phican} +(\partial_x\Laux)^2\Bigr]
\ee
up to corrections of order $e^{-2\phican}$; here $\mu$ is a UV regulator scale for the coincident point singularities of the quadratic terms in functional derivatives
$\delta^2\SS/\delta\phican^2$ and $\delta^2\SS/\delta\phican\delta\Laux$.
The large $\phican$ asymptotic of this wavefunctional agrees nicely with that of the minisuperspace wavefunction, eq.~\pref{Hasymp}.  Since the effects of the Liouville momentum $\varepsilon$ are subdominant at large volume, one expects that just as the leading asymptotic~\pref{Hasymp} of the Hankel function is independent of~$\varepsilon$, this wavefunction should similarly approximate reasonably well the full wavefunction for any Liouville momentum, if the volume is sufficiently large.

The fact that the WKB wavefunctional is independent of $\partial_x\phican$ at large scale factor agrees nicely with the analysis of perturbations in section~\ref {sec:nonlinearity}, in particular eq.~\pref{dphinonlinear}.  Once a given fluctuation of $\phican$ passes outside the horizon, the spatial gradient term in this equation becomes irrelevant compared to the time derivative term and the nonderivative terms due to the rapid growth of the background $e^{2\phib}$, and so it should not be surprising that fluctuations of the spatial gradient of $\phican$ are unsuppressed in the large $\phican$ wavefunction.

%%%%%%%%%%%%%%%%%%%%%%%%%%%%%%%%%%%%%%%%%%
%%%%%%%%%%%%%%%%%%%%%%%%%%%%%%%%%%%%%%%%%%

\subsection{\Back transformation}

In order to gain further control over Liouville theory at the nonlinear level, we need some tools.  One such tool is the \Back transformation, a nonlinear canonical transformation which maps pure Liouville theory on a flat background ($\hat{g}_{ab}=\eta_{ab}$, $\hat{\R}=0$) to a free field theory \cite{Braaten:1982fr,Braaten:1982yn}. 

In the minisuperspace approximation, the integral transform in equation~\pref{Hankelintrep}
\be
\Psi(\phican) = \int d\psi \, e^{iW(\phican,\psi)}\, \Psit(\psi)
~~,\qquad
\Psit(\psi) = \int d\phican \, e^{iW(\phican,\psi)}\, \Psi(\phican) ~~
\ee
with the kernel $W(\phican,\psi)=e^\phican\cosh\psi$, maps the Liouville quantum mechanics Hamiltonian $\frac12\partial_\phican^2+e^{2\phican}$ to the free Hamiltonian $\frac12\partial_\psi^2$ and vice versa; and maps the Liouville eigenfunctions $\Psi(\phican)=H_{i \varepsilon}^{(1)}(e^\phican)$ to plane waves $\Psit(\psi)=e^{-i\varepsilon\psi}$, and vice versa.   

This property has a remarkable generalization to the full 2d field theory.  The free field $\psi$ is related to the Liouville field $\phi$ by the \Back equations
\be
\label{Beq}
\partial_t\phi=\partial_x\psi+\sqrt{\Lambda}e^{\phi}\cosh\psi
~~,\quad 
\partial_x\phi=\partial_t\psi+\sqrt{\Lambda}e^{\phi}\sinh\psi~.
\ee
These equations determine a canonical transformation
\be\label{canonW}
\partial_t\phi=\frac{\delta W}{\delta\phi}
~~,\quad
\partial_t\psi=-\frac{\delta W}{\delta\psi}~,
\ee
with generating functional
\be\label{Backkernel}
W[\phi,\psi]=\int dx \(\phi\partial_x\psi+\sqrt{\Lambda}e^{\phi}\cosh\psi\)~.
\ee
Differentiation of the \Back equations~\pref{Beq} implies $\hat{\nabla}^2 \psi=0$, and
\be\label{TLeq}
{\hat\nabla}^2 \phi+\Lambda e^{2\phi}=0 ~.
\ee
This is the timelike Liouville equation of the action~\pref{STL} on a flat background. Our conventions for the \Back transformation are chosen to make $\phi$ and $\psi$ real for the cosmological solutions ($\Lambda>0$) of interest. To compare with \cite{Braaten:1982yn}, note that the timelike and spacelike Liouville equations in conformal gauge are related by $\Lambda\rightarrow -\Lambda$.

%In fact, the same is true if one replaces both $\Lambda\rightarrow -\Lambda$ and $\sinh\frac{\gamma\psi}{2}\leftrightarrow\cosh\frac{\gamma\psi}{2}$ in~\pref{Beq}. Hence there is a freedom in the \Back equations and a corresponding freedom in $W[\phi,\psi]$. The choice~\pref{Beq} has the advantage that both $\phi$ and $\psi$ are real for the cosmological solutions ($\Lambda>0$) of interest.

For the exact classical solution~\pref{Lexact}, the \Back transformation~\pref{Beq} leads to
\be
\label{Bclassical}
\psi=\frac{1}{2}\ln(-\partial_+{A}/\partial_-{B})~.
\ee
The homogeneous backgrounds~\pref{dsmetrics} correspond to $\psi = \varepsilon t$ for the hyperbolic case, and its continuation $\varepsilon\to 0$ or $\varepsilon \to i\varepsilon$ for the parabolic and elliptic cases.
The quantization of the Liouville theory in terms of the modes of the \Back field was pursued from a somewhat different perspective in~\cite{Gervais:1981gs,Gervais:1982nw,Gervais:1982yf,Gervais:1983am}.

%%%%%%%%%%%%%%%%%%%%%%%%%%%%%%%%%%%%%%%%%%
%%%%%%%%%%%%%%%%%%%%%%%%%%%%%%%%%%%%%%%%%%

\subsection{Ground state wavefunctional}

Consider the \Sch picture wavefunctional for $\psi(x)$ at time $t$, denoted by $\tilde{\Psi}[\psi](t)$. It formally satisfies the \Sch equation
\be
i\frac{\partial}{\partial t}\Psit[\psi]=H_{\psi}\tilde{\Psi}[\psi]
\ee
with free timelike Hamiltonian
\be
\label{HB}
H_{\psi}=\int \frac{dx}{2\pi}\(\frac{\gamma^2}{4}\frac{\delta^2}{\delta\psi^2} - i\partial_x \frac{\delta}{\delta\psi} - \frac{1}{\gamma^2}(\partial_x\psi)^{2}\)~.
\ee
Comparing to the timelike Liouville Hamiltonian
\be
\label{HL}
H_{\phi}=\int \frac{dx}{2\pi}\(\frac{\gamma^2}{4}\frac{\delta^2}{\delta\phican^2}-  \frac{1}{\gamma^2}\Bigl((\partial_x \phican)^{2}-2\partial_x^2\phican-\Lambda e^{2 \phican}\Bigr)\)~,
\ee
one finds the \Sch equation for the Liouville field, $i\partial_t \Psi[\phi]=H_{\phi}\Psi[\phi]$, is formally solved by
\be
\label{BackPsimap}
\Psi[\phi](t)=\int \DD\psi\, e^{(2i/\gamma^2)W[\phi,\psi]}\tilde{\Psi}[\psi](t)~.
\ee
That is, the wavefunctionals for $\phi$ and $\psi$ are related by a \Back transformation at each time $t$.

For compact spatial sections, the free field $\psi$ and its conjugate momentum $\pi_\psi$ on a spatial slice is expanded as
\bea
\psi(x) &=& \psi_{0} + \frac{\gamma}{\sqrt2} \sum_{k\ne 0}e^{ikx}\psi(k) ~,
\nn\\
\pi_\psi(x) &=& p_\psi + \frac{\sqrt 2}{ \gamma} \sum_{k\ne 0}e^{ikx}\pi_\psi(k) ~,
\eea
with reality conditions $\psi^\dagger(k)=\psi(-k)$, $\pi^\dagger_\psi(k)=\pi_\psi(-k)$.  
%To compare to similar formulae in higher dimensions (\cf~\cite{Polarski:1995jg} for an explicit treatment, for instance), one can think of this as a sum over shells of fixed frequency $\omega = |\kk|$ and then there is a sphere of momenta for that frequency.  In one dimension, the sphere consists of two points $k$ and $-k$, corresponding to left and right movers.
One then defines the mode operators
\be
a(k) = \frac{1}{\sqrt2}\Bigl( \sqrt{\omega_k}\, \psi(k) +\frac{i}{\sqrt{\omega_k}}\, \pi_\psi(k) \Bigr) 
\ee
with $\omega_k=|k|$, which evolve as $a(k,t) = e^{-i\omega_k(t-t_0)}a(k,t_0)$, and obey canonical commutation relations
\be
[a(k,t), a^\dagger(q,t)] = \delta_{k,-q}
\ee
In terms of these mode operators, one has
\be
\psi(k) = \frac{1}{\sqrt{\omega_k\gamma^2}}\Bigl({a(k)+a^\dagger(-k)}\Bigr) 
~~,\qquad 
\pi_\psi(k) = -i\sqrt{\frac{\omega_k\gamma^2}{4}} \Bigl( a(k)-a^\dagger(-k) \Bigr) ~.
\ee
To connect to the standard description of worldsheet string theory, we can split the operators into those for positive and negative $k$; those with $k>0$ are the right-movers, while those with $k<0$ are the left-movers (and $\alpha'=\gamma^2/2$).

The ground state wavefunctional $\Psit_{0}[\psi]$ is just a Gaussian for each nonzero mode, times a plane wave for the zero mode:
\be
\label{gdstate}
\Psit_{0}[\psi](t) = C_0 \,
e^{-i E_0 t -i k_\psi \psi_0}
\prod_{k> 0}e^{-\omega_{k}|\psi_{k}|^{2}/\gamma^2}
\ee
where $C_0$ is a normalization constant, and again $k>0$ corresponds to right-movers, $k<0$ to left-movers.
%\emil{The $\psi_k$ here are the amplitudes at some reference time $t_0$, {\it not} the dynamic $\psi(t)$ which is a combination of position and momentum when referred to the reference time $t_0$.  See Polarski and Starobinsky, gr-qc/9504030.}
The ground state satisfies the time-independent \Sch equation 
\be
H_{\psi}\Psit_{0}[\psi] =  E_{0}\Psit_{0}[\psi]
~~,\qquad
E_{0} =  - \frac{\gamma^2}{4}k_\psi^2 + \frac{1}{\gamma^2}Q_\phican^2 ~.
\ee
The zero-mode momentum $k_\psi$ corresponds to the parameter $\varepsilon$ of the classical homogeneous solutions~\pref{dsmetrics}.

%%%%%%%%%%%%%%%%%%%%%%%%%%%%%%%%%%%%%%%%%%
%%%%%%%%%%%%%%%%%%%%%%%%%%%%%%%%%%%%%%%%%%

\subsection{Semiclassical quintessence}

In section~\ref{sec:ModelEquivalence}, we saw that timelike Liouville was related by a field redefinition to a quintessence model with a soft exponential matter potential.  This raised a puzzle, as the quintessence model is expected to exhibit slow-roll eternal inflation, while the free field matter theory is not.  The nonperturbative semiclassical wavefunction provided by the \Back transformation helps explain what is going on.  Let us apply the field redefinition~\pref{fieldredefn} to the semiclassical ground state wavefunctional.  We take the latter to be given by the \Back transform~\pref{BackPsimap} of the free field wavefunction~\pref{gdstate}, tensored with the matter free field wavefunction for $\inflaton$ (we suppress the spectator free fields needed to put us in the weak-coupling limit of Liouville theory, as they are not important for what follows):
\bea
\label{LXwavefn}
\Psi[\phi,X](t) &=& 
\Psi_{0}[\inflaton](t)\; \Psi_0[\phi](t) 
\nn\\
\Psi_{0}[\inflaton](t) &=& C_\infsub \exp\Bigl[\Bigl(-iE_{\infsub}t - ik_\infsub x_0 - \sum_{k> 0} \frac{\omega_k}{2}|\inflaton_k|^2 \Bigr)\Bigr] 
\\
\Psi_0[\phi](t)  &=& \int \DD\psi\, e^{(2i/\gamma^2)W[\phi,\psi]}\tilde{\Psi}_0[\psi](t) 
\nn\\
\Psit_{0}[\psi](t) &=& C_\psi \exp\biggl[\frac{2}{\gamma^2}\biggl(-iE_{\phi}t - ik_\psi \psi_0 - \sum_{k> 0} \frac{\omega_k}{2}|\psi_k|^2 \biggr)\biggr]  ~.
\nn
\eea
Here $x_0$ is the zero mode of $\inflaton$, and $E_\infsub = k_\infsub^2 + \frac 14Q_\infsub^2$.
%\emil{again need to check conventions on zero mode energies.}
The dependence of the combined wavefunction on coordinate time disappears when the zero mode of the Hamiltonian constraint is imposed; this condition (which in string theory terms is the mass-shell condition) sets
\be
-\frac{1}{\gamma^2}k_\psi^2 + k_\infsub^2 + k_\perp^2 = \frac14 \Bigl( \frac{4}{\gamma^2} Q_\phican^2 - Q_\infsub^2 \Bigr) 
\ee
where $k_\perp^2$ is the contribution of free matter fields other than the (field-redefined) inflaton.
As written, the wavefunctions for the zero modes are wavelike and delocalized, but we may take superpositions that localize the fields as suitably localized wavepackets.  Now consider the field redefinition~\pref{fieldredefn}.  The wavefunctionals are now localized separately in
\bea
\phi &=& + \tilde\phi\, \cosh\boost  - \frac\gamma2 \tilde\inflaton \,\sinh\boost  
\nn\\
\inflaton &=&  - \frac2\gamma \tilde\phi \,\sinh\boost  + \tilde\inflaton \,\cosh\boost ~.
\eea
These results help explain what has happened to the supposed dynamics of slow-roll eternal inflation.  Suppose the wavepacket for the free field $X$ is gaussian localized near $X=0$.  Then fluctuations away from $  \frac2\gamma \tilde\phi \,\sinh\boost  \sim \tilde\inflaton \,\cosh\boost $ are gaussian suppressed.
The standard logic of eternal inflation posits that the inflaton fluctuates on its potential, and then the back-reaction of gravity on the fluctuated matter favors a large growth of the scale factor in parts of the wavefunction where the field has fluctuated up the potential over other possible outcomes.  However, what we see in the Liouville-quintessence model is that a fluctuation of the inflaton $\tilde\inflaton$ up its potential is inextricably correlated with a fluctuation of the scale factor $\tilde\phican$ to {\it smaller} values -- you can't have one without the other, because $\inflaton$ is gaussian localized.  Slow-roll eternal inflation requires the geometry and inflaton to be independent actors, whose fluctuations feed one another to generate a sort of runaway behavior; in the Liouville-quintessence model, they are not independent, and there is no runaway.  

Configurations with the inflaton located further up the potential at a given value of the scale factor, or larger scale factor for a given point on the potential, are in the dominant part of the probability distribution for a state built with $X$ localized around a different value.  One must be careful about making such an assertion -- two-dimensional scalars don't have expectation values~\cite{Hohenberg:1967zz,Mermin:1966fe,Coleman:1973ci}.  However, this property refers to the delocalization of massless scalars in 2d due to long-time infrared wandering of the field, or to observations on extremely short time scales.  Here we are interested in what is happening to the matter field $X$ over finite time scales, in finite volume; we can localize the field in a wavepacket, which will undergo quantum spreading, but only by a finite amount in the finite conformal time it takes the scale factor to reach infinite spatial volume.  

While the center of mass will not wander appreciably over finite worldsheet conformal times, the width of the matter wavepacket might be considered to be spreading substantially, depending on what is deemed the appropriate cutoff in the theory or what is an appropriate observable.  For instance, we may wish to consider the spread in $X$ at a fixed proper distance $\Delta = e^\phican dx$; since $e^\phican \sim 1/t$, we will have $dx\sim t$ and thus $\langle X(x) X(x+\Delta)\rangle \sim \log(t)$.  Nevertheless, these effects are not anything that the dynamics is responding to, and moreover they don't push the $X$ field in the preferred direction that would be required by slow-roll eternal inflation.

The standard slow-roll eternal inflation logic requires there to be two independent actors interacting with one another -- the inflaton and the geometry -- when in fact there is only one gauge invariant degree of freedom.  This was already apparent at the linearized level, where the quadratic action depends only on the Mukhanov-Sasaki variable $v$.  The accounting of symmetries guarantees this result.  The geometry has four scalar degrees of freedom, both in the modified Liouville gravity and in four-dimensional Einstein gravity; the inflaton makes a fifth scalar.  There are two scalar gauge invariances, local time reparametrizations and the longitudinal component of spatial reparametrizations.  These two invariances permit us to make two gauge choices, and enforce two constraints.  These four restrictions on the configuration space leave one physical field theoretic degree of freedom out of the original five, not just at the linearized level but in the full nonlinear theory.  In the Liouville-quintessence model, there appears to be sufficient control over the full nonlinear theory to conclude that there is no runaway instability of the sort predicted by slow-roll eternal inflation.  

An important actor here is the Hamiltonian constraint, which ties the configuration of the scale factor to that of the inflaton.  Before it is imposed, the two are independent dynamical fields; one could entertain the notion that the fluctuations of one drive the other.  One should not (as is often done in semiclassical treatments of gravity) impose the Hamiltonian constraint in some averaged way, having gravity back-react independently and classically on the expectation value (or on some tail of the probability distribution) of the quantum-fluctuating matter fields.  Instead, one should impose the Hamiltonian constraint in the fully quantum theory of both matter and geometry.  The prediction of slow-roll eternal inflation is that a matter field in the quintessence frame jumps up its potential, by having gravity classically back-react on the quantum state of matter.   In the Liouville-quintessence model as viewed in the pure \dS frame, this dynamics corresponds to a preferred direction of motion of a free matter field on its flat potential.  Classically, there is no such motion, and the additional quantum motion represents additional stress energy that has not been accounted for in applying the Hamiltonian constraint.  The Hamiltonian constraint forbids the sort of field dynamics that would allow slow-roll eternal inflation to proceed. 

%%%%%%%%%%%%%%%%%%%%%%%%%%%%%%%%%%%%%%%%%%
%%%%%%%%%%%%%%%%%%%%%%%%%%%%%%%%%%%%%%%%%%

\subsection{Excited states}

Matter excitations above the homogeneous \dS background must be properly gravitationally dressed, so that the state continues to satisfy the gauge constraints.  In string theory, the physical, transverse oscillation modes $X^i$, $i=1,...,d-2$ of the string can be used to construct physical, gauge invariant states when appropriately dressed by the timelike and longitudinal modes $X^0$, $X^{d-1}$, and the Faddeev-Popov ghosts $b,c$ in conformal gauge.  The idea is that any given excited state built on the ground state of zero-mode momentum $k^\mu$ by exciting the oscillator modes of the $X^i$, $X^{d-1}$ whose polarizations are transverse to the spatial momentum $\vec k$, and satisfying the mass shell condition
\be
\label{Imsoexcited}
-k_0^2 + \vec k^2 + \NN+\tilde\NN - \frac14 \Bigl(Q_0^2 - \vec Q^2 \Bigr)  = 0
\ee
as well as $\NN=\tilde\NN$ for the total oscillator excitation levels $\NN$ and $\tilde\NN$ of left- and right-movers, respectively, characterizes a representative of a distinct BRST cohomology class.  One can find another representative of that same class which satisfies the Virasoro highest weight condition for the $X^\mu$, and has no Faddeev-Popov ghost excitations.  This characterization of the physical state space holds outside the critical dimension $d=26$ provided one suitably improves one or more of the free field stress tensors of the $X^\mu$ as in eq.~\pref{Timproved}~\cite{Bouwknegt:1991yg,Bilal:1992sn,Benedict:1996qy}.

In the application to 2d \dS cosmology, the modified Liouville theory provides a pair of scalars $\phican$, $\Laux$ which play the role of the timelike and longitudinal coordinates; and $X^i$ are the matter fields.  More precisely, let $\phiperp$ be the field space direction orthogonal to $\phican$ in the gravity sector (\ie\ $\phiperp = \phican+\Laux$ is a free field, decoupled from the Liouville field $\phican$ in conformal gauge).  One then identifies the timelike coordinate $\phican$ with $X^0$, and the spacelike coordinate $\phiperp$ with $X^{d-1}$.  In the \Back representation, $\psi$ plays the role of $X^0$.  Implicitly the construction of elements of BRST cohomology is a form of gravitational dressing of the matter excitations -- we can apply an arbitrary collection of the creation operators of the matter fields to a ground state, solve the mass shell condition~\pref{Imsoexcited}, and the result will be a physical state up to BRST trivial contributions.  There will be some equivalent state with additional excitations of $\phiperp$ and $\psi$ that satisfies the positive frequency half of the Hamiltonian and momentum constraints without Faddeev-Popov ghost excitations.  The \Back transform of this state is a physical excited state of matter in 2d \dS spacetime, as a functional of the matter fields $X^i$ and the gravitational degrees of freedom $\phican$, $\phiperp$ or equivalently $\phican$, $\Laux$.

One can adapt a bit of early string theory technology~-- the DDF operators~\cite{DelGiudice:1971fp}~-- to make the construction of physical states in the modified Liouville theory a bit more explicit.  Let $X^i(t,x)$ be free massless scalar matter fields.  The creation and annihilation operators appearing in the mode sum
\be
X^i = x_0^i + 2\pi p^i\, t + \sum_{n\ne 0} \frac{1}{\sqrt{2n}}\Bigl( a_n^i e^{in(t+x)} + \tilde a_n^i e^{in(t-x)} \Bigr)
\ee
are not gauge invariant, so if one applies them to the oscillator vacuum one will not obtain a physical state.  Consider instead the operators
\be
A_n^i = \oint \frac{ dx}{2\pi} \partial_+ X^i e^{in\kappa X^+}
~~,\qquad
\tilde A_n^i = \oint \frac{ dx}{2\pi} \partial_- X^i e^{in\kappa X^+}
\ee
where $X^+$ is another free field parametrizing a null direction in the (Minkowski) field space metric which is orthogonal to those parametrized by the $X^i$.  These operators commute with the constraints and have a canonical commutation algebra among themselves,
\be
[ A_n^{i}, A_m^{j} ] = m\delta^{ij} \delta_{m,-n} \kappa p^+ 
~~,\qquad
[ \tilde A_n^{i}, \tilde A_m^{j} ] = m\delta^{ij} \delta_{m,-n} \kappa p^+ ~.
\ee
Acting on a ground state of zero-mode momentum $q$, the operators $A_n^i$, $\tilde A_n^i$ are single-valued if we fix $\kappa = 1/q^-$.   
Here $X^-$ is the null direction conjugate to $X^+$ and $q^-$ is the corresponding zero-mode momentum.  These operators thus provide a gauge-invariant version of the  commutation relations of the transverse oscillator modes.  Modified Liouville theory changes this story slightly, due to the light-like linear dilaton.  One must ensure that the DDF operators still commute with the constraints; this will still be true if we identify $X^+$ with the null direction $\Laux$ (or its analogue $\phiperp-\psi$ in the \Back representation).

A coherent state of the DDF operators $A_n^i$, $\tilde A_n^i$ is a semiclassical excited state.
In the familiar case of a single harmonic oscillator of frequency $\omega$, a coherent state is labelled by a complex number $\alpha$ and is built from the vacuum by the displacement operator $D(\alpha)$:
\be
|\alpha\rangle=D(\alpha)|0\rangle,\qquad{\,}D(\alpha)=e^{\alpha\had-\alpha^{*}\ha}~.
\ee
It evolves under the harmonic oscillator Hamiltonian as
\be
|\alpha,t\rangle=e^{-i\omega t/2}|\alpha(t)\rangle,\qquad{\,}\alpha(t)=\alpha e^{-i\omega t}~,
\ee
%\emil{N.B.: in the 2d field theory application, the zero point energies are renormalized away; the residue in the effective theory is the $R \frac1\Delta R$ term, which is the gravitational response to the matter fluctuation determinant.  See Polchinski's discussion of `conformal normal ordering'.}
and its wavefunction is a minimal uncertainty wavepacket
\be\label{psialpha}
\psi_{\alpha}(x,t)=e^{i\xdc(t)\[x-\xc(t)/2\]} \psi_{0}\big(x-\xc(t),t\big)~,
\ee
where $\psi_{0}(x,t)$ is the Gaussian ground state wavefunction. The expectation value of the position operator is the classical trajectory
\be
\xc(t)=\sqrt{\frac{2}{\omega}}\, \mathrm{Re}\bigl(\alpha(t)\bigr)~.
\ee
Coherent states are non-orthogonal and their mutual overlaps are time-independent,
\be
\left\langle{\psi_{\alpha}} | {\psi_{\beta}}\right\rangle=e^{-\half|\alpha|^{2}-\half|\beta|^{2}+\alpha^{*}\beta}~,   
\ee
with the overlap falling rapidly for $|\alpha-\beta|^2\gtrsim 1$.

A coherent state of the gravitationally dressed free fields $X^i(t,x)$ is defined in complete analogy,
\be
\exp \[ \sum_{n>0} \( \alpha_n^i A_n^i - {\alpha_n^{* i}} A_{-n}^i  + \tilde\alpha_n^i \tilde A_n^i - {\tilde\alpha}_n^{* i} \tilde A_{-n}^i \) \]  \ket{\Psit_0} ~.
\ee
We thus expect a coherent state wavefunctional that is a localized, minimal uncertainty wavepacket evolving in the physical time $X^+$, or ultimately the \Back field $\psi$, with overlaps of these states given by (rewriting $\tilde\alpha^i_n \equiv \alpha^i_{-n}$)
\be
\langle \Psit_\alpha | \Psit_\beta \rangle = \exp \Bigl[ - \sum_{n\ne 0} \frac n2\Bigl( |\alpha_n|^2 + |\beta_n|^2 - 2 \alpha_n^*\beta_n \Bigr) \Bigr] ~,
\ee
independent of $X^+$.
The \dS space wavefunctionals in terms of $\phican$, $\Laux$ are given again by the \Back transform of the wavefunctional of $\psi$, and the longitudinal free field $\xi$ replaced by $\phican+\Laux$.  The Liouville-quintessence wavefunctionals are given by the further field redefinition~\pref{fieldredefn}, where $X$ is one of the transverse matter fields $X^i$.

%%%%%%%%%%%%%%%%%%%%%%%%%%%%%%%%%%%%%%%%%%%%%%%
%%%%%%%%%%%%%%%%%%%%%%%%%%%%%%%%%%%%%%%%%%%%%%%

\subsection{To infinity...and beyond!?}

In \dS geometries, the scale factor blows up at finite conformal time coordinate, just as it blows up at finite conformal spatial coordinate in anti-deSitter space.  In the latter case, there is a natural conformal boundary condition for Liouville theory~-- the Zamolodchikovs' ZZ boundary state~\cite{Zamolodchikov:2001ah}.  The boundary state formalism in 2d conformal field theory characterizes the boundary conditions on conformal fields at a finite conformal boundary; for instance they characterize D-branes in string theory.  The ZZ boundary state is roughly equivalent to a Dirichlet boundary condition that sets $\phi=\infty$ at the conformal boundary.  In the presence of the ZZ boundary, correlation functions scale as the locations of operators approach the conformal boundary in the manner that one expects in two-dimensional AdS space; in particular, $e^{2\phi}$ has all the properties of the AdS scale factor.  This result led one of the authors to propose in~\cite{Martinec:2003ka} that the natural description of conformal infinity in two-dimensional \dS space would be the analytic continuation of the ZZ boundary state to timelike Liouville theory.

The characterizing property of conformal boundary states is that they impose reflecting boundary conditions on the stress-energy tensor, 
\be
\label{bdystate}
T_{++} = T_{--} ~.
\ee
In classical Liouville theory, the stress tensor is the Schwartzian derivative of the free fields $A$ and $B$:
\bea
T_{++} &=& \frac{1}{\gamma^2}\[ \frac{\partial_+^3 A}{\partial_+ A} - \frac32 \(\frac{\partial^2_+ A}{\partial_+ A}\)^{\!\! 2} \]
\nonumber\\
T_{--} &=& \frac{1}{\gamma^2}\[ \frac{\partial_-^3 B}{\partial_- B} - \frac32 \(\frac{\partial^2_- B}{\partial_- B}\)^{\!\! 2} \]  ~.
\eea
It was shown in~\cite{Gervais:1981gs} that two functions that have the same Schwarzian derivative are related by a constant M\"obius transformation
\be
B(x) = \frac{a A(x) + b}{c A(x) + d}~~,\qquad ad-bc=1 ~.
\ee
For the conformal boundary we are interested in, if we place the boundary at $t=0$, the boundary condition is $B(x-t))|_{t=0}=A(x+t)|_{t=0}=f(x)$, \ie\ $a=d=1$, $b=c=0$; this determines $\phi=\infty$, and also sets $\psi=0$ at the boundary.  Thus, at the semiclassical level, the \dS Liouville boundary state seems to be an ordinary D-brane boundary state for the \Back field $\psi$.  Formally, the Liouville boundary state wavefunctional would then be given by the \Back transformation of the free field Dirichlet boundary state for $\psi$.

If the gravity state of the system at $\phi=\infty$ is a Liouville boundary state, that state should be tensored with a boundary state of the matter fields and Faddeev-Popov ghosts.  This leads to a puzzle: The boundary condition for free matter fields is also something of the same sort, for example a Dirichlet or Neumann boundary state.  Such a boundary state reflects left-moving modes into right-moving modes.  This implies a halving of the number of independent states one can define on a Cauchy surface, since the left-moving and right-moving initial data have to be correlated in such a way as to satisfy the reflecting boundary condition in the future.  The puzzles here seem to be of much the same sort as those that arise in the `black hole final state' proposal of~\cite{Horowitz:2003he}.
{In the black hole context, the final state boundary condition must accomplish a similar feat, correlating the infalling matter that made the black hole with the negative energy partners of the Hawking quanta, which fall into the singularity.}  
In the cosmological context, it would seem much more sensible to allow arbitrary initial data, but then the classical solution moves the conformal boundary to another location, see for example equation~\pref{bdyloc}.  One will also need some suitable modification of the matter boundary conditions if the initial data is allowed to be arbitrary.  The gauge constraints should correlate the matter initial data to the Liouville initial data, and the matter and Liouville boundary states.  In other words, the `final state boundary condition' in such a scenario must have some sort of initial state dependence if we demand the initial data allow independent values to be set for left- and right-movers.

These issues are arising because the notion of a Penrose diagram for \dS space (and indeed, more generally -- for instance in the black hole evaporation problem) is to some extent a fiction, an artifact of the classical approximation.  For classical \dS space, the classical geometry reaches infinite scale factor $\phican=\infty$ at finite conformal time $t$.  In the quantum theory, asking the question `at what coordinate time does the quantum mechanical variable $\phican$ reach the value $\phican_0$' is a meaningless question, and so we should not be able to say with certainty where the conformal boundary is located, if it is defined as the coordinate time when $\phican=\infty$ is reached.

It is also not clear how seriously to take the implicit classical relation~\pref{Lexact}, \pref{Bclassical} between the \Back field $\psi$ and the Liouville field $\phican$, at the quantum level.  For semiclassical wavepackets, the saddle point of the \Back transformation is concentrated at the classical value for each field; however, the integral over $\psi(x)$ in the transform on the third line of~\pref{LXwavefn} runs over all values of $\psi$, not just the negative ones that appear classically for homogeneous \dS space.
So if we let $\psi$ range over all the reals, what is the wavefunction defined by the Backlund transform describing?

%%%%%%%%%%%%%%%%%%%%%%%%%%%%%%%%%%%%%%%%%%
%%%%%%%%%%%%%%%%%%%%%%%%%%%%%%%%%%%%%%%%%%

\subsection{`in-in' formalism}

An intriguing possibility is suggested by an analysis of Mathur~\cite{Mathur:1993tp}.%
\footnote{A rare instance where the answer to a titular question may be `yes'.}
Suppose that, instead of calculating the `in-out' transition amplitude between some initial state and a boundary state at conformal infinity, we were interested in computing some `in-in' amplitudes for a given `in' state.  
According to~\cite{Mathur:1993tp} (see also \eg~\cite{Niemi:1983nf}), the Schwinger-Keldysh formalism~\cite{Schwinger:1960qe,Mahanthappa:1962ex,Mahanthappa:1963a,Mahanthappa:1963b,Keldysh:1964ud} calculates such processes in the first-quantized path integral along a contour that runs from the infinite past to the infinite future, and then back to the infinite past, perhaps after a discrete shift along the imaginary time axis.  Consider now the evolution of the \Back field $\psi$; as $t$ runs over $t\in (-\infty,+\infty)$, $\psi$ evolves from $-\infty$ to $+\infty$, passing through $\psi=0$ in particular.  Chasing this through the classical \Back transformation, one sees that $e^\phi$ then runs from zero to $\infty$, jumps to $-\infty$, and runs back down to zero; in other words, the trajectory of the timelike field $\phi$ is precisely what one wants for the calculation of `in-in' amplitudes.  
%It is also argued in~\cite{Mathur:1993tp} that the zweibein changes sign when the trajectory starts running backward.  These are all properties of the \dS classical solutions if we extend the coordinate domain beyond the locus where $\phi_\cl$ diverges.  In this way, we neatly sidestep the issue raised above.

Perhaps the simplest examples of this structure are provided by the spatially homogeneous solutions~\pref{dsmetrics}, for example the hyperbolic solution $\sqrt\Lambda e^\phi = -\varepsilon/\sinh[\varepsilon t]$.  Consider this semiclassical trajectory, and define the classical Liouville field through the \Back transformation of the free field $\psi=\varepsilon t$; let $\psi$ propagate freely along an infinite 2d cylindrical geometry.  There is no conformal boundary at finite $t$ in the $\psi$ representation, and thus no confusion about the `cosmological final state'; instead all possible final states in the vicinity of $t=0$ are summed over automatically.  The location of the slice where $\phi=+\infty$ fluctuates in a natural way, and we have an ensemble of late-time states in the cosmology.  At large positive coordinate time $t$ one has a flipped copy of the `in' state, satisfying the other half of the constraints that were not imposed on the original `in' state.

A hint that the integral over all values of the \Back field $\psi$ in a plane wave state corresponds to this sort of path in $\phican$ space is provided by the \Back transform at the level of minisuperspace quantum mechanics.  The integral transform with kernel $e^{iW}$, $W=e^\phican\cosh \psi$, maps a plane wave in $\psi$ to the Hankel wavefunction for Liouville
\be
{i\pi}\,e^{-\pi \varepsilon/2}H_{i\varepsilon}^{(1)}(e^\phican) 
	= \int^\infty_{-\infty} \!d\psi\; \exp\bigl[{i\, e^\phican\cosh \psi}\bigr]
		\,\exp[-i\varepsilon\psi]   
\label{Hankelrep}
\ee
but the inverse transform yields a cosine
\be
\frac{-2i\,e^{\pi \varepsilon/2}}{\varepsilon\,\sinh(\pi \varepsilon)}\;\cos(\varepsilon\psi)
	= \int_{-\infty}^\infty\! {d\phican}\,
		\exp\bigl[{i\, e^\phican\cosh \psi }\bigr]\,
		H_{i\varepsilon}^{(1)}(e^\phican)
%\quad .
\label{Hankelinv}
\ee
(equivalently, the Hankel function $H^{(1)}_{i\varepsilon}(e^\phican)$ and the cosine $\cos(\psi)$ are a dual transform pair on the half-line $\psi<0$).
If one asks what Bessel function integrates to a pure plane wave under the inverse transform with kernel $e^{iW}$, the answer is
\bea
\frac{i\coth(\pi\varepsilon)}{\varepsilon} \exp[-i\varepsilon\psi]
	&=& \int_{-\infty}^\infty\! {d\phican}\,
		\exp\bigl[{i\, e^\phican\cosh \psi }\bigr]\,
		J_{i\varepsilon}(e^\phican)
\label{Besselpair}
\\
&=& \int_{-\infty}^\infty\! {d\phican}\,
		\exp\bigl[{i\, e^\phican\cosh \psi }\bigr]\,
		\frac12\Bigl( H^{(1)}_{i\varepsilon}(e^\phican) + \overline{{H^{(1)}_{i\varepsilon}(e^\phican)}}\, \Bigr)
\nn
\\
&=& \int_{-\infty}^\infty\! {d\phican}\,
		\exp\bigl[{i\, e^\phican\cosh \psi }\bigr]\,
		\frac12\Bigl( H^{(1)}_{i\varepsilon}(e^\phican) - e^{-\pi\varepsilon}{H^{(1)}_{i\varepsilon}(-e^\phican)} \Bigr) ~.
\nn
\eea
In other words, the plane wave state in $\psi$ is the \Back transform of the Hankel function along a contour that runs up the real $\phican$ axis to infinity and then back after a shift by $i\pi$.  In the full 2d field theory, the interpretation of the plane wave state of the \Back field may involve a similar path in $\phi$ space.  The analysis of~\cite{Harlow:2011ny} lends support to the notion of using such complex integration contours.

%%%%%%%%%%%%%%%%%%%%%%%%%%%%%%%%%%%%%%%%%%
%%%%%%%%%%%%%%%%%%%%%%%%%%%%%%%%%%%%%%%%%%

\subsection{On the probability of a probability interpretation}

The Hamiltonian constraint is a wave operator on functionals that is second order in functional derivatives.  Thus it has a structure more akin to the Klein-Gordon equation than the \Sch equation.  As a consequence, there is no natural positive definite inner product on the space of solutions; the best one can do is a Klein-Gordon style norm,
but it is not clear what form such a norm would take in the present context.  In string theory in asymptotically flat spacetime, one imagines defining such a norm for each of the gauge-fixed mass eigenstates of the string.  In the string theory interpretation of timelike Liouville theory, this structure holds in the far past (small scale factor $e^\phican$), but in the far future (large scale factor $e^\phican$) the behavior becomes a bit wilder.  The `normal modes' of $\phican$ are in some sense the free left- and right-moving modes of $A(x^+)$, $B(x^-)$ which also classically define the \Back field $\psi$;
%\footnote{The quantization of Liouville theory in terms of these modes was pursued in~\cite{Gervais:1981gs,Gervais:1982nw,Gervais:1982yf,Gervais:1983am}.}
however, as we saw above, small fluctuations of $A$, $B$ lead to wild excursions of $\phican$ at late times.

The Klein-Gordon norm is not positive definite.  In Klein-Gordon field theory, this property implies that there is in general no probability interpretation at the level of single particle states -- one must pass to a field-theoretic description.  The probability interpretation is at the level of the Fock space of multiparticle (rather, `multi-universe') states.  As mentioned in section~\ref{sec:objections}, the pair production rate in this class of `tachyon' backgrounds diverges~\cite{Strominger:2003fn}.  This leads to a `measure problem' which seems not all that different, at least in spirit, to that encountered in eternal inflation -- the universe one wants to study in the ensemble of all universes has measure zero in that ensemble.  Nevertheless, the description of that single universe seems to make sense, at least at the semiclassical level; we should also note that the \Back presentation seems not to have such an explosion of pair production occurring.  Furthermore, this proliferation of universes is different from that usually considered in eternal inflation in that there is no drift of the inflaton relative to classical expectations.

A different inner product is given by the two-point function of conformal field theory, which leads to the Zamolodchikov metric on the Hilbert space~\cite{Zamolodchikov:1986gt,LeClair:1988sp,Zwiebach:1992ie}.  This inner product is an integral over all the configuration space, including the timelike coordinate $\phican$, rather than a spacelike slice.  It is not clear what interpretation to give this quantity in the present context given that, as mentioned earlier, the CFT two-point function given by the conformal bootstrap does not seem to be diagonal in the conformal weight~\cite{Zamolodchikov:2005fy,Strominger:2003fn,McElgin:2007ak}.

We leave the question of the appropriate notion of norm, and the probability interpretation, as an interesting topic of further research.

%%%%%%%%%%%%%%%%%%%%%%%%%%%%%%%%%%%%%%%%%%%%%%%
%%%%%%%%%%%%%%%%%%%%%%%%%%%%%%%%%%%%%%%%%%%%%%%

%%%%%%%%%%%%%%%%%%%%%%%%%%%%%%%%%%%%%%%%%%%%%%%
%%%%%%%%%%%%%%%%%%%%%%%%%%%%%%%%%%%%%%%%%%%%%%%

\vskip .5cm
\noindent
%{\bf Acknowledgements:}
\acknowledgments
We thank Guilherme Pimentel for useful discussions, and Peter Adshead and Mark Wyman for comments on the manuscript. This work was supported in part by DOE grants DE-FG02-90ER-40560 and DE-FG02-13ER41958. W.E.M. gratefully acknowledges support from the US Department of State through a Fulbright Science and Technology Award.

%%%%%%%%%%%%%%%%%%%%%%%%%%%%%%%%%%%%%%%%%%%%%%%
%%%%%%%%%%%%%%%%%%%%%%%%%%%%%%%%%%%%%%%%%%%%%%%

\vskip 2cm

%\newpage
%\bibliographystyle{amsunsrt-ensp}
\bibliographystyle{JHEP}
\bibliography{EImodel}

\end{document}